\documentclass[reprint,
superscriptaddress,
amsmath,amssymb,
aps,
prb,
longbibliography
]{revtex4-2}

\usepackage{amsthm,latexsym,amssymb,amsmath,verbatim}
\usepackage{textcomp}
\usepackage{gensymb}
\usepackage{graphicx}
\usepackage{dcolumn}
\usepackage{hyperref}
\usepackage{chemformula}
\usepackage[T1]{fontenc}
\usepackage{longtable}
\usepackage{multirow}
\usepackage{booktabs}
\usepackage{bm}
\usepackage{xcolor}
\usepackage{array}
\usepackage{makecell}

\begin{document}
\title{Dislocation-templated antiferromagnetic domains in epitaxial NiO}

\author{Alexandra Fonseca Montenegro}
\affiliation{Department of Materials Science and Engineering, The Ohio State University, Columbus OH 43210, USA}
\author{Justin Michel}
\affiliation{Department of Physics, The Ohio State University, Columbus OH 43210, USA}
\author{Fengyuan Yang}
\affiliation{Department of Physics, The Ohio State University, Columbus OH 43210, USA}
\author{Roberto C. Myers}
\email{myers.1079@osu.edu}
\affiliation{Department of Materials Science and Engineering, The Ohio State University, Columbus OH 43210, USA}
\affiliation{Department of Physics, The Ohio State University, Columbus OH 43210, USA}
\affiliation{Department of Electrical and Computer Engineering, The Ohio State University, Columbus OH 43210, USA}

\date{\today}

\begin{abstract}
Controlling antiferromagnetic domains is essential for spintronics, yet predictable manipulation remains challenging due to their lack of net magnetization. Here, we utilize scanning electron microscopy electron channeling contrast imaging (ECCI) to observe the formation of nanoscale antiferromagnetic domains in epitaxial NiO/MgO(001) at interface dislocations. Thermal cycling across the N\'eel temperature reveals that while domain contrast vanishes in the paramagnetic phase, the features re-emerge at identical spatial locations upon cooling. Diffraction-vector-dependent ECCI imaging, governed by a relative tilt contrast model, indicates that the domains stem from localized rhombohedral distortions of volumetric twin-type $T$-domains. These $T$-domains are stabilized by the long-range shear stress field of interface misfit dislocations (MDs). By tracking a film thickness series from 23 to 46~nm, we resolve a transition in oxide relaxation mechanics from primary $60^\circ$ interface MDs along $\langle110\rangle$ to a cross-slipped network of pure edge $90^\circ$ MDs deviated toward $\langle100\rangle$. We demonstrate that ECCI identifies the Burgers vector and line direction of these MDs via strict $\vec{g} \cdot (\vec{b} \times \vec{u})$ invisibility criteria, and an Eshelby interaction energy model accurately predicts the asymmetric nucleation of specific $T$-domains along MD flanks. Quantitative image analysis shows $T$-domain volume fractions track the spatial configuration of the evolving MD networks, providing a framework for defect-engineering antiferromagnetic textures using one-dimensional defects.
\end{abstract}

\maketitle

Antiferromagnetic (AFM) materials have emerged as a compelling platform for next-generation spintronics due to their ultra-fast spin dynamics, robustness against stray magnetic fields, and absence of cross-talk in high-density devices.\cite{baltz_antiferromagnetic_2018,jungwirth_antiferromagnetic_2016} However, the very characteristic that makes AFM systems attractive (lack of net magnetization) renders the control, pinning, and patterning of their domain structures an unsolved challenge.\cite{cheong_seeing_2020} Unlike ferromagnets, which can be easily reconfigured via external magnetic fields, AFM domains typically freeze into stochastic layouts governed by random thermal fluctuations during cooling cycles across their N\'eel temperature ($T_N$). Devices relying on the AFM order parameter are limited by the domain structure since the N\'eel vector (spin-orientation) changes direction between AFM domains. For example, spin-transport in AFMs is degraded by the presence of domain walls (DWs) within a spin channel.\cite{lebrun_long-distance_2020, ross_propagation_2020} Developing a scalable method to deterministically pattern AFM domain structure remains a barrier for reliable AFM-based logic and memory devices.

Unlike ferromagnets, perfect antiferromagnets lack a macroscopic stray dipolar field to drive multidomain formation. As originally deduced by N\'eel\cite{Neel1954} and later inferred from macroscopic observations in bulk NiO by Roth\cite{rothNeutronOpticalStudies1960} and Slack\cite{slack_crystallography_1960}, the formation and stabilization of AFM domains are instead intrinsically tied to localized strain fields from crystallographic defects. Recent macroscopic observations of thermal memory in AFM domains for NiO/MgO(001) across $T_N$ strongly support this,\cite{xu_imaging_2019} implying that a fixed underlying defect array pins the domain structure. However, directly establishing the real-space nanoscale defect template responsible for enforcing this memory has remained elusive. To date, resolving these embedded AFM textures has relied primarily on synchrotron-based techniques such as X-ray magnetic linear dichroism photoelectron emission microscopy (XMLD-PEEM)\cite{cheong_seeing_2020, scholl_observation_2000, weberMagnetostrictiveDomainWalls2003a, moriyama_micromagnetic_2023}, or localized scanning probe methods including nitrogen-vacancy (NV) center magnetometry\cite{gross_real-space_2017} and spin-polarized scanning tunneling microscopy\cite{bode_atomic_2006}. While these approaches provide remarkable spatial sensitivity, they are mainly surface-sensitive, probing only the first few monolayers of a material. This prevents the simultaneous probing of both buried localized defect strain and sub-surface domain structures.

In epitaxial rocksalt oxide thin films grown under lattice mismatch, such as NiO on MgO(001), strain relaxation proceeds through the nucleation and propagation of extended one-dimensional defects.\cite{james_thickness-dependent_1999} As these films surpass a critical thickness, threading dislocations (TDs) glide through specific slip systems, leaving behind structured arrays of interface misfit dislocations (MDs).\cite{freundThinFilmMaterials2004} These dislocation lines introduce strong elastic stress fields ($\boldsymbol{\sigma}^{\text{MD}}$) near their cores,\cite{andersonTheoryDislocations2017} breaking the cubic symmetry of the host lattice, generating emergent properties like ferroelectricity and ferromagnetism that are absent in bulk materials.\cite{gao_atomic-scale_2018, sugiyama_ferromagnetic_2013} Because AFM order in NiO is intrinsically coupled to a small spontaneous rhombohedral magnetostrictive shear distortion ($\eta=-0.58\times10^{-3}$) along the four degenerate $\langle111\rangle$ directions below $T_N$,\cite{roth_magnetic_1958,slack_crystallography_1960,yamada_spin_1966,nakahigashi_crystal_1975,mandal_strain-induced_2009} these MD stress fields can break the degeneracy, and stabilize or destabilize $T$-domains depending on the mechanical coupling between $\boldsymbol{\sigma}^{\text{MD}}$ and the magnetoelastic eigenstrain ($\boldsymbol{\varepsilon}^*$) of an AFM domain. Through the Eshelby magnetoelastic interaction, the asymmetric shear stress of the MD lowers the nucleation energy for specific twin domains ($T$-domains), naturally serving as a template for the nanoscale AFM architecture (Sec.~S.6.B). This concept is illustrated in Fig.~1a, showing a $T$-domain forming in a region where the MD stresses stabilize the magneto-elastic strain.
\begin{figure}
\centering
\includegraphics[width=3.33in]{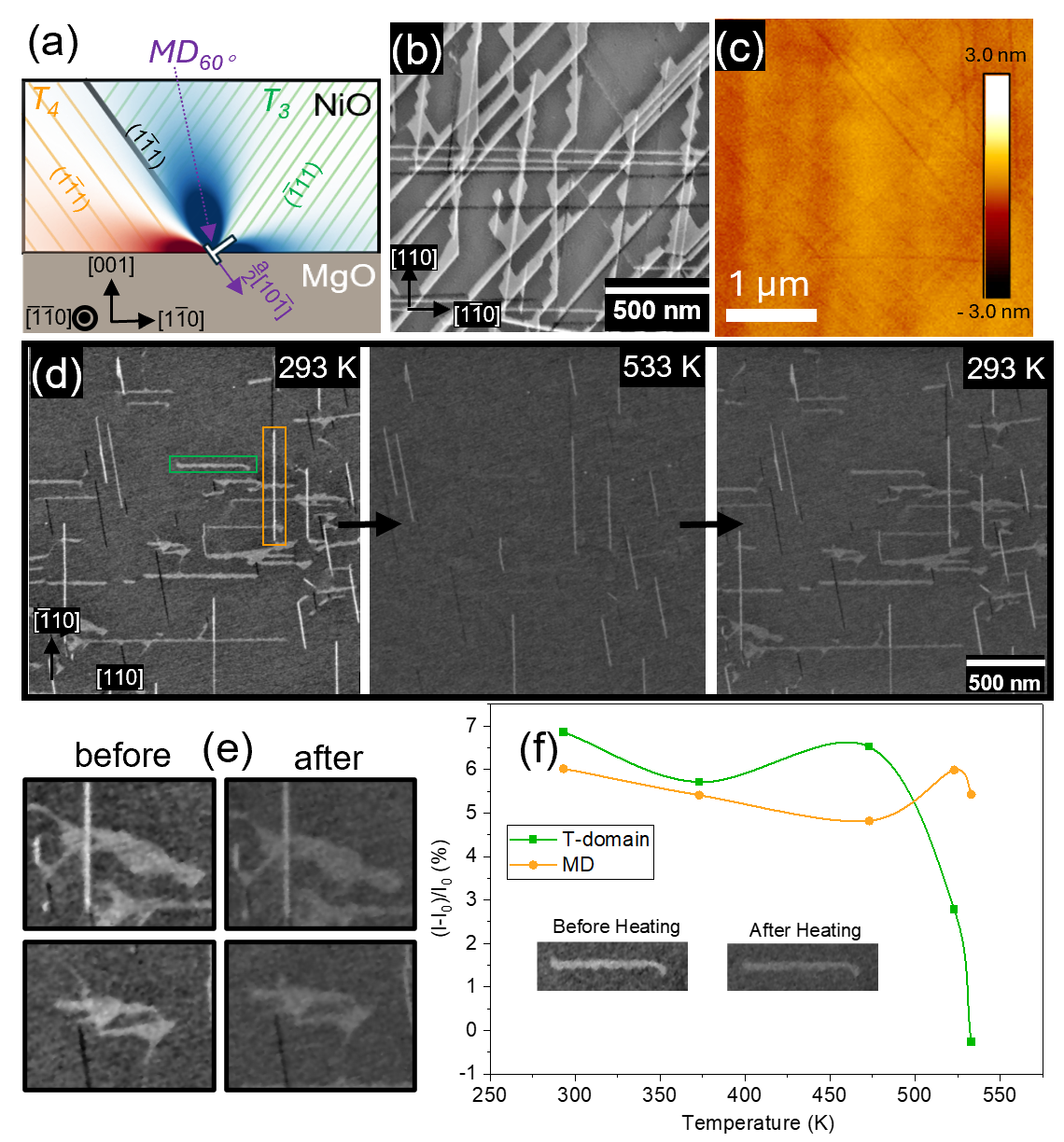}
\caption{(a) Schematic of an antiferromagnetic $T$-domain forming near a misfit dislocation (MD) at a NiO/MgO(001) interface. The shear-stress associated with the $(\bar{1}12)$ extra-half-plane of an MD with $\vec{b}$ (purple) and $\vec{u}=[\bar{1}\bar{1}1]$ breaks the T-domain degeneracy, creating a magnetoelastic energy minimum (blue) that templates a $T_3$ domain composed of alternating up and down Ni-spins coupled ferromagnetically within $(\bar{1}11)$ (green) planes. Below the $(1\bar{1}1)$ slip plane (black), $T_4$ (orange) is stable. (b) ECCI of 33~nm NiO ($\vec{g} = (2\bar{2}0)$) showing MDs and T-domains. (c) AFM topography of the same region. (d) Thermal cycling ECCI across $T_N = 533~\text{K}$. (e) Zoomed view showing precise T-domain recovery. (f) Normalized ECCI contrast versus temperature.}
\label{epitaxy}
\end{figure}

In this Letter, we provide experimental verification of this principle at the nanoscale. We exploit scanning electron microscopy (SEM) electron channeling contrast imaging (ECCI) to simultaneously correlate buried interfacial MD arrays with the overlying rhombohedral $T$-domain strain distributions in epitaxial NiO(001). In-situ thermal cycling across $T_N = 523 \pm 4$~K demonstrates a structural memory effect where $T$-domains located at MDs vanish in the paramagnetic state but re-emerge at identical MD-sites upon cooling. Using quantitative Eshelby energy modeling and diffraction vector ($\vec{g}$) dependent ECCI, we show that the asymmetric shear stress of the MD dictates the stabilization of specific 3D volumetric $T$-domain slabs. Furthermore, we unveil a thickness-driven transition in the plastic relaxation pathways of the rocksalt epilayer. The initial relaxation occurs via $60^\circ$ MDs (lines along $\langle110\rangle$) on $\{111\}$ planes), which act as a kinetic pathway before cross-slipping to form higher strain-relieving pure edge $90^\circ$ MDs (lines along $\langle100\rangle$) on $\{110\}$ planes). Quantitative image analysis reveals that the volume fractions and spatial orientations of the $T$-domains tightly track this evolving MD configuration, establishing a framework for defect-engineering stable AFM textures.

Epitaxial NiO films of varying thickness ($23$--$60$~nm) were grown on MgO(001) substrates by RF magnetron sputtering under optimized oxygen partial pressure (Sec.~S.1). Structural characterization and macroscale strain relaxation were assessed using high-resolution X-ray diffraction (HRXRD) indicating smooth interfaces and high crystallinity for all films. Due to the propensity of NiO vacancy formation, on-axis HRXRD is insufficient to determine the relaxation state, requiring reciprocal space mapping  (RSM) (Fig.~S1.b). Assuming the Poisson ratio 0.21, for the thickest 60~nm film the NiO is 11\% relaxed from a maximum tensile mismatch of +0.64\% to the MgO substrate. Unlike the globally averaged RSM data, ECCI provides a microscale image of individual strain fields, capable of tracking discrete dislocations\cite{boyer_correlation_2021, haidet_versatile_2023,yan_multi-microscopy_2024, fonseca_montenegro_log-normal_2024} and local structural domains\cite{ihlefeld_domain_2017,peng_electron_2026}. ECCI was performed in a field-emission SEM equipped with a backscattered electron detector. Imaging conditions were tuned to isolate crystallographic defects and domains by selecting specific diffraction vectors ($\vec{g}$) based on experimental Electron Channeling Patterns (ECP). Temperature-dependent ECCI was carried out using an in-situ heating stage (Sec.~S.2).

Representative images of dislocation and AFM domain networks are presented in Fig. \ref{epitaxy}. At room temperature, ECCI of a 33~nm NiO film reveals the co-existence of sharp line segments from interfacial MDs and broader patchy contrast regions (Fig. \ref{epitaxy}b). Both features show a systematic $\vec{g}$-dependence confirming their strain-induced origins (Figs. 2, S4--S6, and S9). Atomic force microscopy (Fig. \ref{epitaxy}c) confirms a uniform step-terrace morphology, ruling out topographic artifacts. 

During in-situ thermal cycling, the sample was heated to 533~K, 10~K above the bulk reported $T_N$ (523~K), to ensure the magnetic phase transition was fully crossed (Fig.~1d). In this paramagnetic phase, the broad patchy features vanish while the MD lines remain. Upon cooling back to 293~K, the domains re-emerge at identical spatial coordinates and with the same geometry. Quantitative spatial correlation of registered ECCI micrographs yields a Pearson coefficient of $R = 0.79$ within the domain terrace and a $>96\%$ binarized area match across the field of view (Sec.~S.3). This indicates that immobile MDs serve as the underlying crystallographic template responsible for the AFM structural memory first reported by Xu et al.\cite{xu_imaging_2019}, providing a direct nanoscale view of defect-driven domain formation originally deduced by N\'eel\cite{Neel1954}, Roth, and Slack\cite{rothNeutronOpticalStudies1960, slack_crystallography_1960}.
\begin{figure}
\includegraphics[width=3.33in]{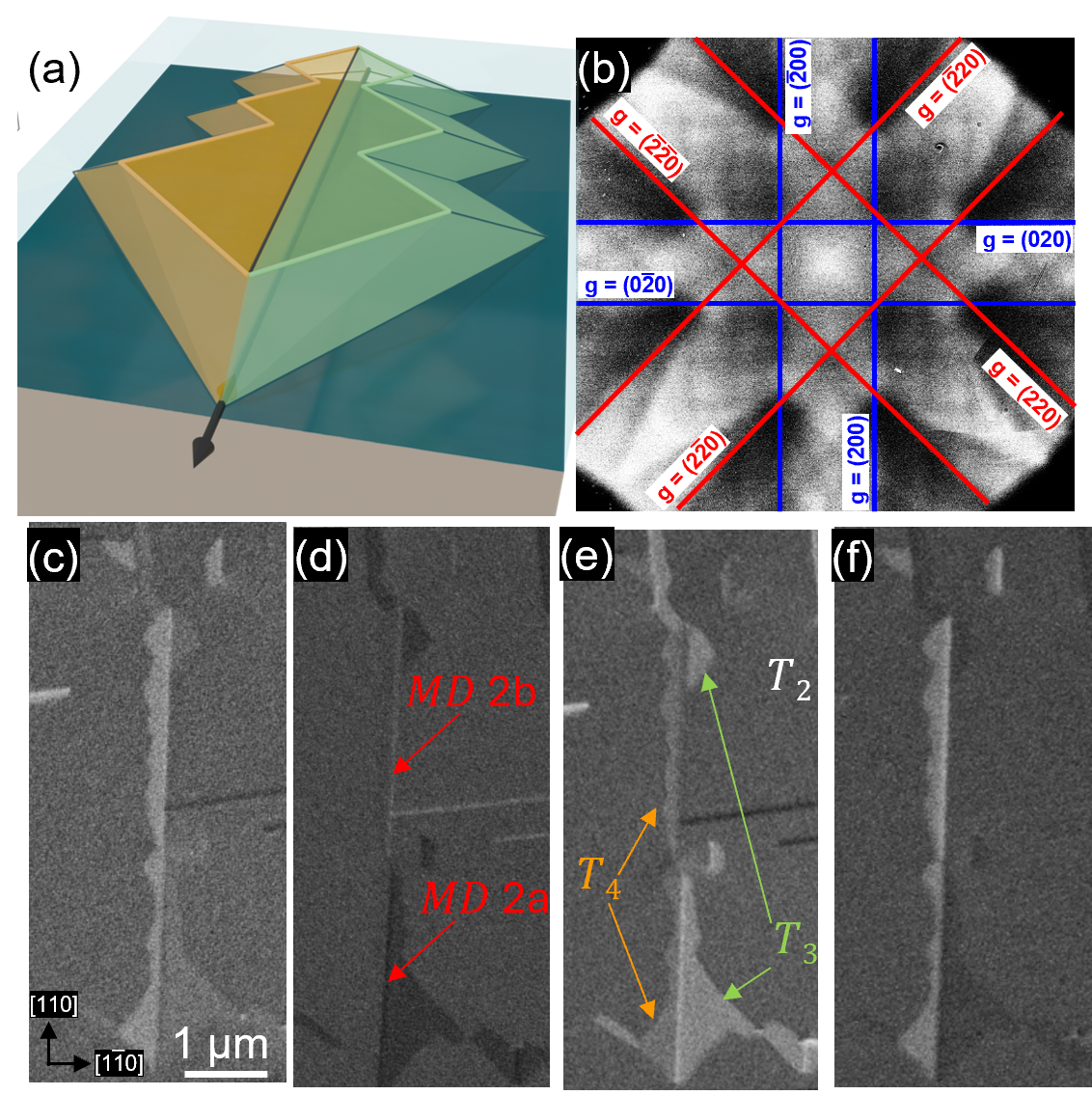}
\caption{(a) Schematic 3D projection of a $T_3$ and $T_4$ domain flanking opposite sides of an MD (black arrow) in a NiO film with $T_p=T_2$. (b) Low-magnification ECP showing selected $\vec{g}$-conditions. ECCI images of 28~nm NiO across 4 conditions: (c) $\vec{g_1} = (200)$,  (d) $\vec{g_2} = (0\bar{2}0)$, (e) $\vec{g_3} = (220)$ and (f) $\vec{g_4} = (2\bar{2}0)$, enabling unique identification of the visible $T_4$ and $T_3$ flanking domains, and the invisible surrounding $T_2$ parent domain (see Secs.~S5--S6 of the Supplementary Material).}
\label{ECCI}
\end{figure}

To determine the nature of the domain structures, ECCI images were recorded across four distinct in-plane $\vec{g}$ (Fig.~2(c)-(d)). Visible domains display uniform, flat intensity profiles with sharp edges ($36\pm3$~nm) (Fig.~S6), indicating we are imaging 3D volumetric $T$-domain (\textit{easy-plane}) slabs rather than wide bulk domain walls that have reported widths of $<100$~nm,\cite{simonnot_atomic-scale_2026,weberMagnetostrictiveDomainWalls2003a} which would appear in ECCI as continuous contrast gradients instead of the observed uniform plateaus. Strain due to $S$-domains are not expected to be visible, as this \textit{easy-axis} magnetostriction ($\lambda<10^{-5}$) is at least $5\times$ weaker\cite{nakamichiAntiferromagneticMagnetostrictionCoO1961b} than $\eta$, and below the sensitivity limit of ECCI\cite{picardTheoryDynamicalElectron2014} (Sec.~S.5.A).  

When a secondary $T$-domain is embedded within a parent ($T_p$) domain, the difference in their spontaneous rhombohedral strain tensors relaxes into a relative out-of-plane lattice tilt vector $\boldsymbol{\omega}^{(k,p)}$ (Sec.~S.5.B). The resulting ECCI contrast shift is directly proportional to the projection of this tilt onto the diffraction vector, $C \propto \mathbf{g} \cdot \boldsymbol{\omega}^{(k,p)}$. In Fig.\ref{ECCI}(c-f), two domain variants (labeled $d_2$ and $d_3$ in the Supplementary Material) are observed on eithe side of MDs. Their ECCI contrast sequence identifies them as $T_4$ and $T_3$ domains with $T_2=T_p$ (Table~S3). For example, the ECCI contrast for $T_4$ is completely extinct in Fig. \ref{ECCI}(b), $\vec{g_2} = (0\bar{2}0)$, and nearly vanishes for  $T_3$ in Fig. \ref{ECCI}(f), $\vec{g_4} = (2\bar{2}0)$. The boundary between these flanking domains is along the MD $[\bar{1}\bar{1}0]$ line, which forms the interface edge of a vertical $(1\bar{1}0)$ plane. Roth predicted this plane as one of the two possible $T_3/T_4$ domain walls,\cite{rothNeutronOpticalStudies1960} the other being (001), which being parallel  to the film is geometrically impossible. The boundary of the domains with the parent (away from the MD), shows a faceted structure that is explained by the possible (habit planes) domain walls. According to Roth, the  $T_2/T_4$ walls are $(10\bar{1})$ or (010), while the $T_2/T_3$ walls are $(0\bar{1}1)$ or (100).\cite{rothNeutronOpticalStudies1960} Therefore, the projection of these domain walls onto the [001] interface would consist of alternating $\langle100\rangle$ in-plane boundaries, 45$^{\circ}$ from the MD $[\bar{1}\bar{1}0]$ line, e.g. $T_2/T_4$ alternating between $(10\bar{1})$ canted, or (010) vertical walls. This is illustrated in Fig. \ref{ECCI}(a), which illustrates a proposed 3D nanoscale $T$-domain structure consistent with the ECCI images and twin-boundary crystallographic restrictions. 

\begin{figure}
\includegraphics[width=2.5in]{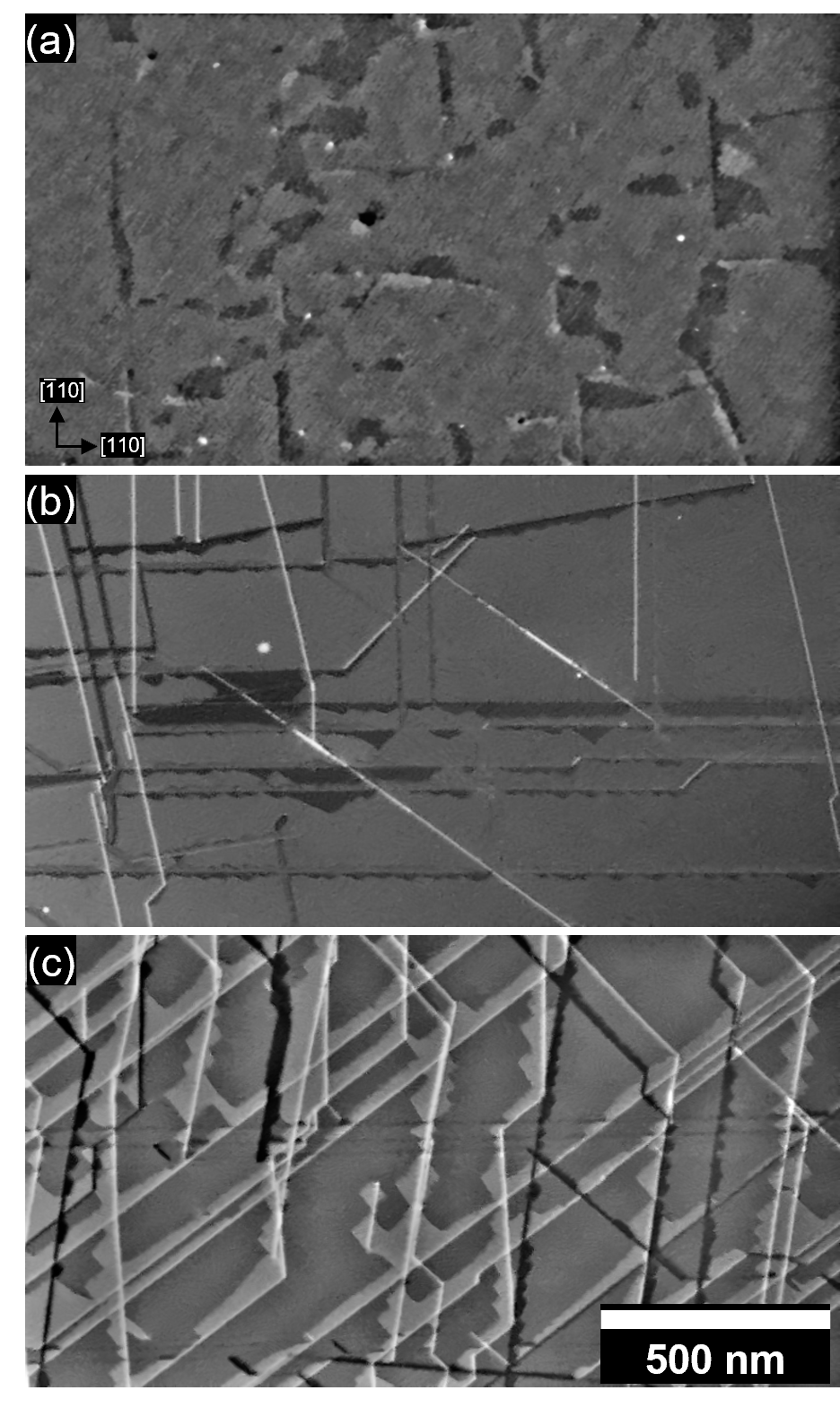}
\caption{Thickness-dependent ECCI showing a transition in oxide relaxation. (a) 25~nm NiO dominated by sparse primary $60^\circ$ MDs. (b) 33~nm NiO showing the onset of cross-slipped pure-edge $90^\circ$ MDs. (c) 46~nm NiO displaying a dense, cross-slipped network with T-domains flanking the MDs.}
\end{figure}

\begin{figure}
\centering
\includegraphics[width=3.33in]{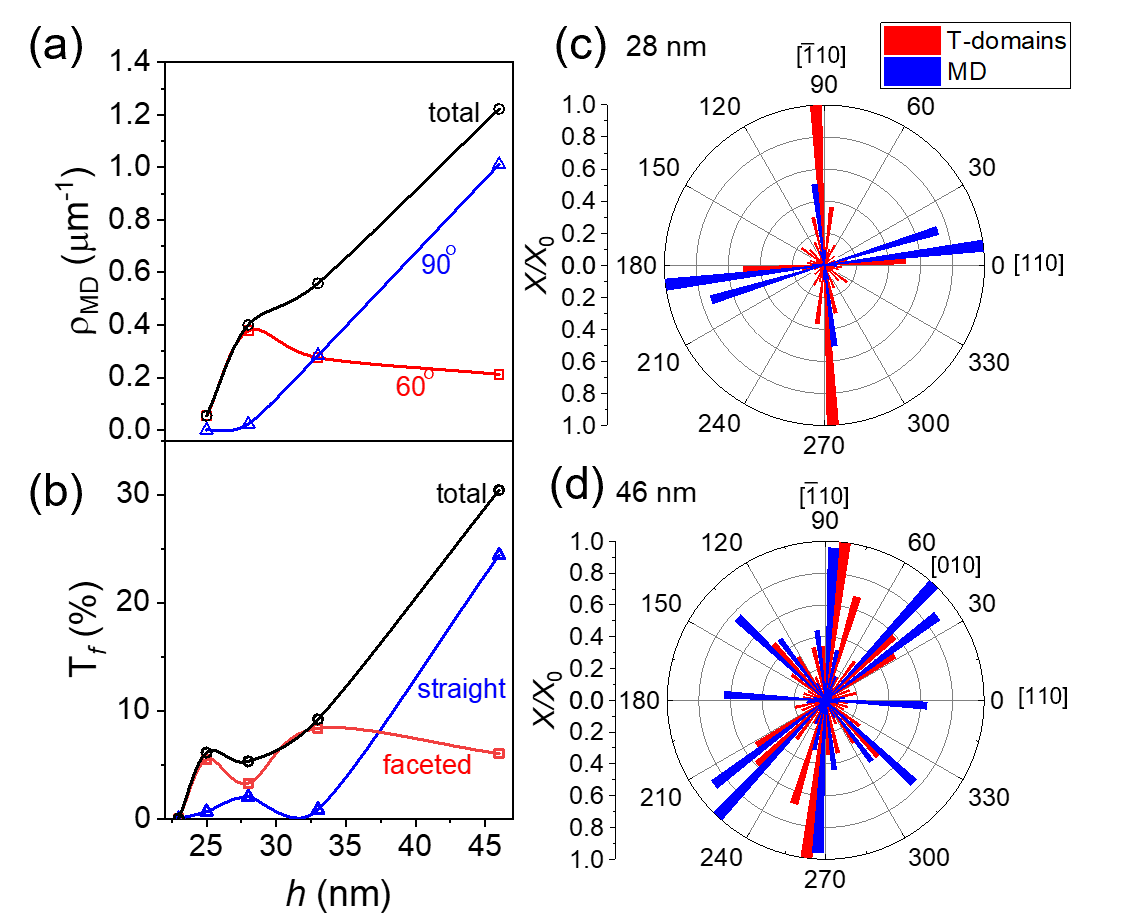}
\caption{(a) Linear MD density ($\rho_{MD}$) versus film thickness ($h$) mapping the cross-slip transition from $60^\circ$ to $90^\circ$ dominant MD populations. (b) Corresponding T-domain area fraction ($T_f$) shift from faceted to straight domain walls. (c--d) Angular distribution histograms correlating MD line directions and $T$-domain long axes for 28~nm and 46~nm films.}
\label{stats}
\end{figure}

ECCI imaging of MDs as a function of film thickness and $\vec{g}$ allows determination of the slip systems active during strain-relaxation in NiO epilayers. We found that both $60^\circ$ and $90^\circ$ MDs strictly obey a $\vec{g} \cdot (\vec{b} \times \vec{u})$ visibility criterion, permitting the unique assignment of the Burgers vector $\vec{b}$ and line direction $\vec{u}$ (Sec.~S.4.C, Table~S2). A fundamental transition in the rocksalt epilayer relaxation mechanics is illustrated in Fig.~3, where initial plastic relaxation at 25~nm occurs via $60^\circ$ MDs gliding on primary $\{111\}$ planes, but by 33~nm and 46~nm, TDs cross-slip (Sec.~S.4.B) onto secondary $\{110\}$ planes to maximize strain relief, forming a dense network of long, pure edge (Lomer) $90^\circ$ MDs. While a pure-edge interfacial segment alone does not prove pure screw TDs, this explicit observation of cross-slipped MD segments shifting from $60^\circ$ (primary) to $90^\circ$ (secondary), combined with the strict requirement of $\vec{b}$-conservation, dictates the presence of pure screw TDs. Similar MDs are reported in (001) PbSe epilayers.\cite{haidet_versatile_2023} The SEM-ECCI measured MD linear density (line length per area) $\rho_{MD}$ is plotted in Fig. \ref{stats}(a) as a function of film thickness for the two MD line directions, showing an effective slip-system cross-over around 33~nm.

The correspondence between the MD networks and the templated $T$-domains is illustrated in Fig. \ref{stats}, where $\rho_{MD}$ and $T$-domain area fraction $T_f$ closely track together for each MD/T-domain pairing. Specifically, we observe the density of child T-domains with faceted domain walls (e.g., $T_3/T_2$ and $T_4/T_2$ in Fig. \ref{ECCI}) following that of the $60^\circ$ MDs. In contrast, the density of child domains forming straight walls with $T_p$ (e.g., Fig. S6) follow the density of $90^\circ$ MDs. As summarized in the polar plots in Fig. \ref{stats}(c) and (d), the long-axis of the $T$-domains generally tracks the line direction of the underlying MDs. The local angular offsets observed are attributed to the complex topology of the domains; unlike 1D dislocations, the domains facet and merge at MD intersections, which skews the raw long-axis angular distributions.

The thermodynamic stabilization of specific $T$-domains at these MDs is explained by a local magnetoelastic interaction. Using an Eshelby formalism (Sec.~S.6), we map the mechanical work done by the MD stress field during the spontaneous rhombohedral pure shear transformation (eigenstrain $\boldsymbol{\varepsilon}^*$). Because the interaction energy scales as $E_{\text{int}} \propto -\sigma_{yz}^{\text{MD}} \varepsilon_{yz}^*$, a $T$-domain variant with a matching shear eigenstrain experiences a deep energy minimum on one specific flank of the MD core, facing a positive energy barrier on the opposite side (Sec. S.6.B). An example is shown in Fig. \ref{epitaxy}a, for a $T_3$ domain with $(\bar{1}11)$ easy-planes (green) interacting with an MD with line vector $\vec{u}=[\bar{1}\bar{1}0]$, Burgers vector $\vec{b}=\frac{a}{2}[10\bar{1}]$ (purple), and slip plane $(1\bar{1}1)$. For this 60$^{\circ}$ MD (labeled MD 2a in Supplementary Material), the edge component is characterized by an atomic half-plane $(\bar{1}12)$, which generates a large compressive shear stress ($\sigma_{yz}$). Since it closely aligns with the compressive $T_3$ shear strain ($\varepsilon_{yz}^*$), a magnetoelastic energy  minimum (blue shading) occurs above and on the $[1\bar{1}0]$ side (right) of the MD core, in the vicinity of the half-plane. Below the slip-plane, on the opposite flank (left), this domain experiences an energy maximum (red shading), i.e. the tensile shear stress region of the core. $E_{\text{int}}$ is approximately $-120$~meV/nm$^3$ in the extreme minimum. Even at a distance of 3 nm from the core, well outside the limits where continuum elasticity breaks down, the energy minimum remains roughly $-10$~meV/nm$^3$. These energies are orders of magnitude greater than the bulk magnetic anisotropies (Sec.~S.6.C), fully dominating the thermodynamic landscape near the core and dictating domain formation. These intense magnetoelastic interactions likely alter the N\'eel ordering temperature immediately adjacent to the defect line. We evaluated the  $E_{\text{int}}$ profiles (as in Fig. \ref{epitaxy}(a)) for each of the 4 $T$-domains near the 12 possible MDs, 48 possible MD/T-domain combinations (Fig.~S8), and identified the stabilized domain type and its location at each MD (Table~S4). For every unique SEM-ECCI distinguishable $T$-domain observed, the predicted MD-flanking location matches the prediction (Sec.~S.6.D).

In summary, SEM-ECCI provides a direct, non-destructive probe uniquely capable of resolving both buried interface dislocation networks and their overlying volumetric $T$-domain strain distributions. We provide the physical mechanism for robust macroscopic thermal memory effects in NiO AFM films by demonstrating that domain recovery is directly templated by the underlying immobile misfit dislocations. We introduced an ECCI contrast model based on relative lattice tilt and an Eshelby magnetoelastic energy model that together confirm the unique identity of both the MDs and the $T$-domains. The enormous asymmetric shear stress of the dislocation core effectively breaks $T$-domain degeneracy, offering a potential pathway toward the spatial control of domain architectures in antiferromagnetic materials.

\section*{Supplementary Material}
See the supplementary material for comprehensive details on thin film growth, HR-XRD, quantitative image analysis methodologies, and SEM-ECCI characterization and analysis. It contains derivations for $\vec{g}$-dependent ECCI contrast and the Eshelby-based magnetoelastic $E_{\text{int}}$ model for MD/$T$-domain interactions, accompanied by 7 additional data figures and 2 multipanel color maps of $E_{\text{int}}$ near MD cores. The 4 supplementary tables provide: (1) complete crystallographic parameters for each of the 12 possible MDs (slip planes, $\vec{u}$, and $\vec{b}$), (2) the 8 possible cross-slip MD pairs and their unique ECCI contrast sequences, (3) the predicted ECCI sequences for the 12 parent/child $T$-domain combinations, and (4) a tabulation of all 48 MD/$T$-domain combinations, identification of the predicted $T$-domains formed at each MD and flank direction, and example $T$-domain assignments in 2 NiO films.

\begin{acknowledgments}
Financial support from the Air Force Office of Scientific Research (Grant FA9550-23-1-0330, Program Managers Drs. Ali Sayir and Jiwei Lu) and the Department of Energy (DOE), Office of Science, Basic Energy Sciences, under Grant No. DE-SC0001304 (sample growth and XRD characterization) are acknowledged.
\end{acknowledgments}

\newcommand{\myCustomSuppCommand}{...} 

\setcounter{equation}{0}
\setcounter{figure}{0}
\setcounter{table}{0}

\newcommand{\m}[1]{\bar{#1}}

\renewcommand{\thesection}{S.\arabic{section}}
\renewcommand{\thetable}{S\arabic{table}}
\renewcommand{\thefigure}{S\arabic{figure}}

\hypersetup{
    colorlinks=true,
    linkcolor=blue,
    filecolor=magenta,      
    urlcolor=cyan,
}

\clearpage
\onecolumngrid
\section*{Supplementary Material}

\tableofcontents
\clearpage 

\section{Thin film growth and HR-XRD}

Epitaxial NiO films of varying thickness ($20\text{--}60$~nm) were grown on MgO(001) substrates by RF magnetron sputtering under optimized oxygen partial pressure to ensure stoichiometry. Film growth rates were monitored in situ using a quartz crystal microbalance (QCM) with a high-purity ($99.99\%$) Ni target, and total film thicknesses were calibrated via X-ray reflectivity (XRR).

\begin{figure}[htbp]
\centering
\includegraphics[width=6in]{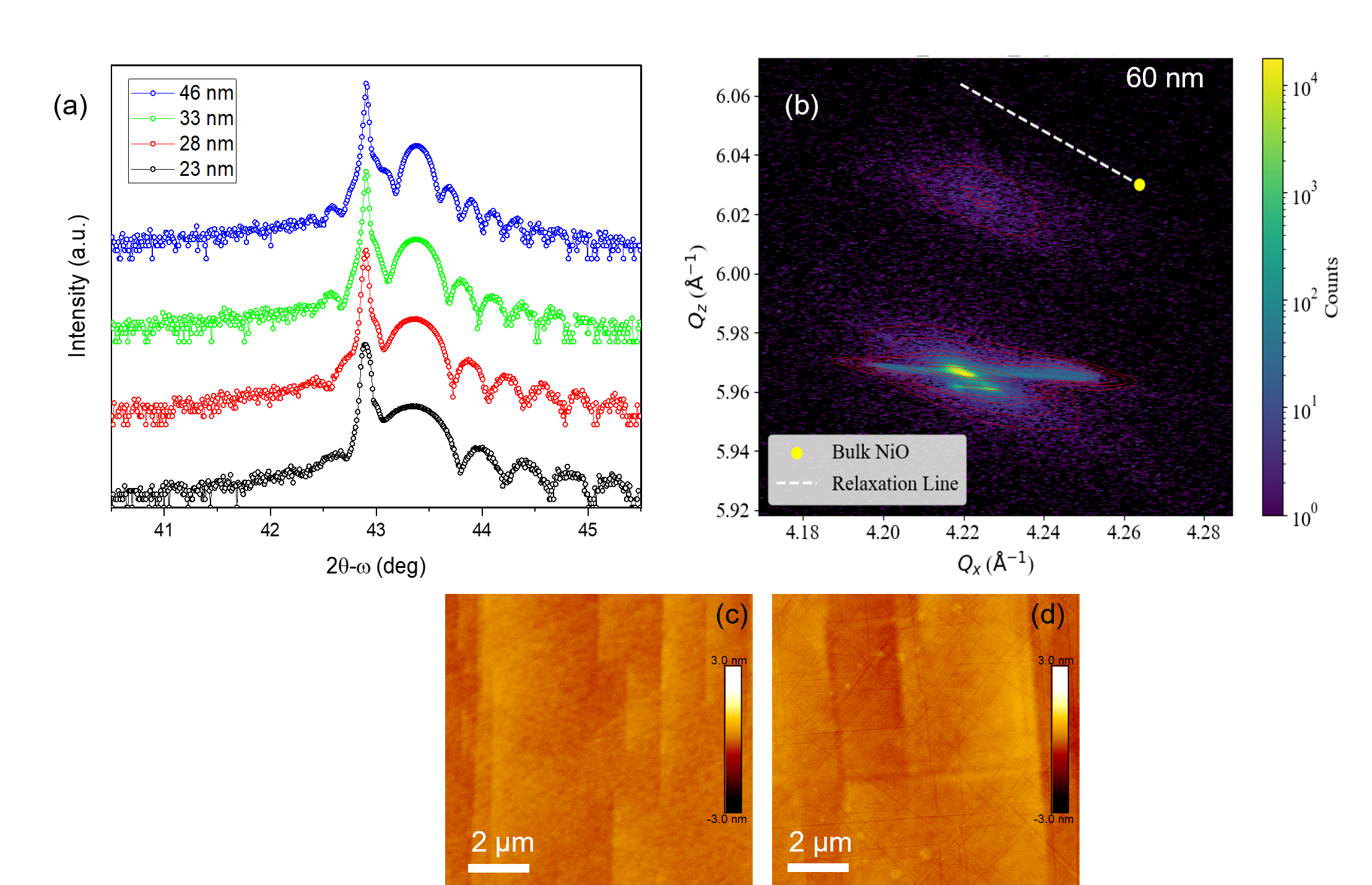}
\caption{Structural and morphological characterization of NiO thin films grown on MgO(001). (a) High-resolution X-ray diffraction (HRXRD) $\omega-2\theta$ scans exhibit well-defined thickness fringes, indicating smooth interfaces and high crystallinity across the series of film thicknesses. (b) Reciprocal space mapping (RSM) of a 60~nm film around the NiO (224) reflection illustrates the degree of lattice relaxation. Atomic force microscopy (AFM) images of (c) a 23~nm and (d) a 46~nm NiO film show comparable step-terrace morphology, demonstrating that surface topography remains largely unchanged with increasing thickness.}
\label{HRXRD}
\end{figure}

HR-XRD and RSM measurements were carried out using a high-resolution X-ray diffractometer equipped with a Cu $K\alpha_1$ source ($\lambda = 1.5406$~\AA) and a 2D area detector. Example HRXRD and RSM data are plotted in Fig.~\ref{HRXRD}. On-axis HRXRD measurements, sensitive to out-of-plane $c$ only, are insufficient to accurately determine the relaxation state of the films due to the propensity of NiO to form vacancies, requiring RSM to measure in-plane $a$ and out-of-plane $c$ independently. For the RSM shown, we find in-plane $a_{\text{NiO}}=4.2073$~\AA\ and $a_{\text{MgO}}=4.2112$~\AA, while out-of-plane $c_{\text{NiO}}=4.1707$~\AA\ and $c_{\text{MgO}}=4.212$~\AA. Thus, the substrate matches the expected lattice parameter for unstrained MgO, while the film exhibits in-plane tension and out-of-plane compression, as expected. 

Assuming the Poisson ratio of NiO, $\nu = 0.21$, the unstrained cubic lattice parameter $a_0$ of NiO is determined, assuming pure biaxial strain, as $a_0=(c(1-\nu)+2\nu a)/(1+\nu) = 4.183$~\AA, which is within values previously reported, yielding a total theoretical mismatch to the substrate of $(4.2112-4.183)/(4.183)=0.64\%$. The measured in-plane strain is $(4.2073-4.183)/(4.183)=0.57\%$, which indicates a degree of relaxation of this film of $1-0.57/0.64 = 11\%$.

\begin{figure}[htbp]
\centering
\includegraphics[width=6.6in]{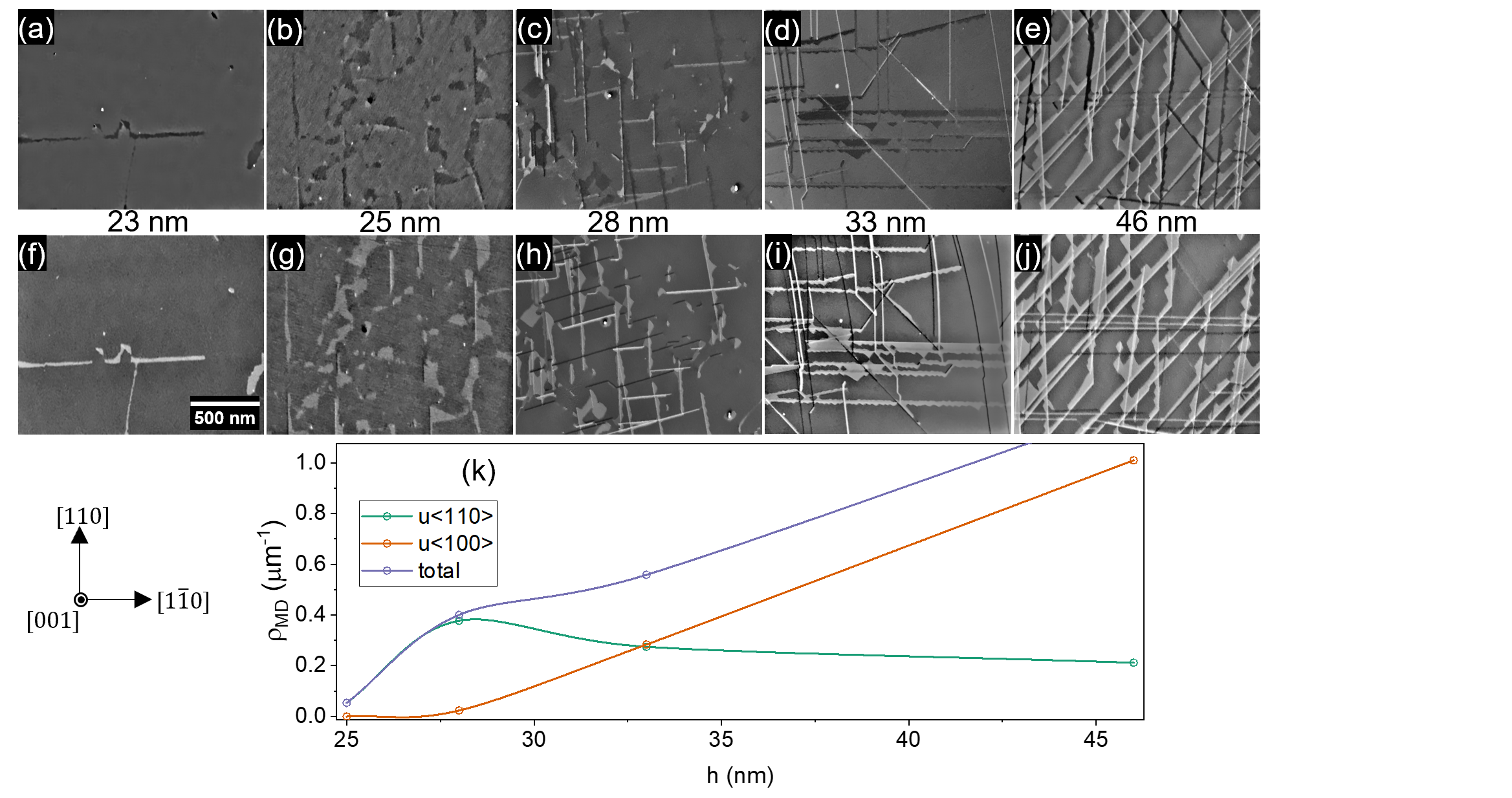}
\caption{ECCI images of NiO (001) epilayers on MgO (001) obtained at $g_3=(220)$ (a--e) and $g_4=(2\bar{2}0)$ (f--j). The thickness of the film is listed between the images. (k) MD linear density ($\rho_{\text{MD}}$) as a function of film thickness ($h$), showing a crossover from initial MD segments along $\vec{u} \parallel \langle110\rangle$ transitioning to $\langle100\rangle$ by cross slip.}
\label{ECCI-thickness}
\end{figure}

Compared with HRXRD or RSM, SEM-based ECCI measurements can characterize very low degrees of relaxation through quantifying the strain-relieving misfit dislocation (MD) content. Example ECCI data from several NiO film thicknesses are presented in Fig.~\ref{ECCI-thickness}, showing an increased density of MDs with film thickness as well as a change in the dominant MD line direction $\vec{u}$ from $\langle110\rangle$ to $\langle100\rangle$ via cross slip. Quantitative image analysis was performed to measure the MD linear density (total length per area), $\rho_{\text{MD}}$, as a function of film thickness ($h$), plotted in Fig.~\ref{ECCI-thickness}(k) (see also main text Fig.~4). In the case of ZnS on GaP (with a similar mismatch of $0.56\%$), a $\rho_{\text{MD}} \approx 3~\mu\text{m}^{-1}$ provides $\approx 10\%$ relaxation for $60^\circ$ MDs of the $\{111\}\langle 110 \rangle$ slip system \cite{fonseca_montenegro_log-normal_2024}. This implies the RSM results in Fig.~\ref{HRXRD} could be accounted for by a similar $\rho_{\text{MD}}$. In Fig.~\ref{ECCI-thickness}(k), we observe a trend of increasing $\rho_{\text{MD}}$ with $h$, at $\approx 1~\mu\text{m}^{-1}$ for $h=46$~nm, which extrapolates to the implied density of $3~\mu\text{m}^{-1}$ for the 60~nm film based on its RSM-measured degree of relaxation.

\section{SEM-ECCI Methods and Analysis}

Electron channeling contrast imaging (ECCI) was performed using a Thermo Fisher QUATTRO environmental scanning electron microscope (ESEM) at the Center for Electron Microscopy and Analysis (CEMAS). Standard ECCI was conducted at an accelerating voltage of 20 kV, a beam current of 2.0 nA, a working distance of 10 mm, and a tilt of $-7^\circ$. To observe the system's behavior across the N\'eel temperature ($T_N$), in-situ thermal cycling was performed using a heating stage. Because the required elevated temperatures exceeded the operational specifications of the standard annular backscattered electron (BSE) detector, a BSE detector diode (K.E. Developments Limited, Type Q, 18 mm) was manually installed below the pole piece. The diode current was collected under zero bias and amplified using an EBIC Amplifier (Deben Type-31) and imaged using the SEM external input to successfully capture ECCI data during the high-temperature phases. To ensure the sample crossed the magnetic phase transition, the film was heated to 533 K, 10 K above the reported bulk \$T\_N\$ of 523 K. This provided a wide safety margin against any thickness-dependent or epitaxial deviations in the film's specific transition temperature.

Crystallographic alignment was precisely calibrated using electron channeling patterns (ECP). Unlike zinc blende, in a perfect MgO rock salt crystal, the $[110]$ and $[1\bar{1}0]$ directions are crystallographically indistinguishable. Therefore, for each individual NiO/MgO specimen, the ECP was used to identify the $\langle 110 \rangle$ and $\langle 100 \rangle$ poles, and a scratch was made on the substrate to define a specific $[010]$ reference axis, ensuring consistent measurement orientation.

Quantitative image analysis of misfit dislocation (MD) linear densities ($\rho_{\text{MD}}$) and $T$-domain area fractions ($T_f$) was conducted using standardized SEM-ECCI acquisition protocols. Micrographs across all epilayer thicknesses ($h = 23, 25, 28, 33, 46\text{ nm}$) were collected at a constant magnification of $8000\times$. For each film, images were acquired across $N = 3$ non-overlapping regions under four orthogonal diffraction conditions ($\vec{g}_1 = 220$, $\vec{g}_2 = \bar{2}\bar{2}0$, $\vec{g}_3 = 2\bar{2}0$, and $\vec{g}_4 = \bar{2}20$). Each individual field of view (FOV) spans $26.0\ \mu\text{m} \times 18.4\ \mu\text{m}$, corresponding to a calibrated sampling area of $478.8\ \mu\text{m}^2$ at a digital resolution of $237\text{ pixels}/\mu\text{m}$.

Image post-processing and feature extraction were executed using MIPAR software. ECCI micrographs were first normalized using background subtraction and histogram equalization to correct for localized illumination gradients. Dislocation line features and $T$-domain areas were segmented via adaptive thresholding, followed by manual morphological inspection. The total integrated line length of the segmented MDs was divided by the FOV area to establish $\rho_{\text{MD}}$, while the total segmented domain area was used to calculate $T_f$. 

To obtain the total cumulative quantities reported in Fig.~4 ($T_f = \sum A_{\text{domain}} / A_{\text{total}}$ and $\rho_{\text{MD}} = \sum L_{\text{MD}} / A_{\text{total}}$), exhaustive statistical sampling was performed over the total analyzed area ($1436.4\ \mu\text{m}^2$ per thickness condition). The total number of $T$-domains measured was 200, 174, 305, 228, and 27 for the 46, 33, 28, 25, and 23 nm films, respectively. The corresponding total numbers of measured MDs were 138, 82, 133, 33, and 0 (none observed). Because the reported metrics are cumulative totals over the standardized area rather than ensemble averages, standard deviations are not defined for these values. Systematic uncertainty is instead governed by the digital spatial resolution ($4.2\text{ nm}/\text{pixel}$) and segmentation boundary definitions. High-resolution line intensity profiles taken across isolated features yield a Gaussian full-width at half-maximum (FWHM) of $33 \pm 5\text{ nm}$ for MD line contrast and an edge transition width of $36 \pm 3\text{ nm}$ for $T$-domain boundaries. 

For the quantified data presented in Fig.~4(c) and (d), both $\rho_{\text{MD}}$ and $T_f$ are plotted within the same polar histogram by normalizing their values against their respective maximums. The raw (unnormalized) $\rho_{\text{MD}}$ and $T_f$ are plotted in Fig.~4(a) and (b), respectively.

\section{Quantification of Domain Thermal Memory}

\begin{figure}[htbp]
\centering
\includegraphics[width=6.6in]{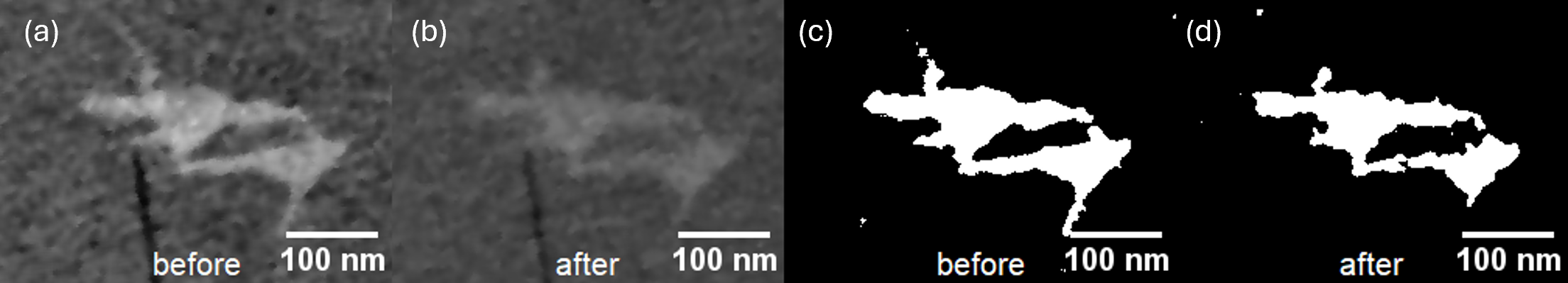}
\caption{ECCI images of a domain structure at room temperature, (a) before, and (b) after cylcing the temperature above and below $T_N$. (c) and (d) show the same images in (a) and (b), respectively, after applying a threshold to filter out dislocation contrast. Pixel-wise XOR calculation of (c) and (d) shows an average mismatch of $<4\%$.}
\label{memory}
\end{figure}

To quantitatively evaluate the thermal memory effect, ECCI micrographs acquired before and after thermal cycling through $T_N$ were spatially registered to account for thermal drift. Registration was performed using the BigWarp plugin in FIJI,\cite{bogovicRobustRegistrationCalcium2016,schindelin_fiji_2012} applying a landmark-based thin-plate spline (TPS) transformation. The stationary terminations and intersections of the misfit dislocation (MD) network were utilized as fiducial markers to ensure precise spatial alignment. Spatial cross-correlation of the unthresholded, registered full-field ECCI data, shown in Fig. 1(d), yielded a strong Pearson correlation coefficient of $R = 0.75$. To confirm that this correlation is driven by the deterministic return of the $T$-domains rather than the static structural contrast of the MDs, the analysis was repeated within a local region of interest (ROI) explicitly chosen to isolate the domain patches from the bright MD cores, Fig. \ref{memory} (a) and (b). Within this cleaner domain ROI, the Pearson correlation increased to $R = 0.79$, confirming robust intrinsic thermal memory within the domain structure itself. To further quantify this memory spatially, the full-field images were binarized to isolate the discrete $T$-domain states, Fig. \ref{memory}(c) and (d). A pixel-wise Exclusive OR (XOR) calculation between the pre- and post-cycling masks yielded an area mismatch of only $3.65\%$, representing a $>96\%$ perfect spatial agreement of the domain configurations. Together, these metrics verify a highly deterministic return to the original domain state upon cooling below $T_N$.

\section{\label{sec:MD-ECCI} Diffraction ($\vec{g}$) Dependent SEM-ECCI Analysis of Misfit Dislocations in NiO (001) Epilayers on MgO (001)}

Spontaneous plastic strain relaxation is mediated by the glide of threading dislocations (TDs) that form interfacial misfit dislocations (MDs) at the film-substrate interface with surface normal $\hat{n}_s = [001]$. In the early stages of relaxation, only $\langle110\rangle$ MD lines are observed, consistent with MDs formed by TD glide within the $\{111\}\langle110\rangle$ slip system. At increased thickness, staircase MD segments are observed with initial MDs changing direction by $45^\circ$, tracing $\langle100\rangle$ segments. These are the line directions $\vec{u}$ expected for MD segments formed by TD glide in the $\{110\}\langle110\rangle$ slip system.

Under the Finish-Start Right-Hand (FS/RH) convention, the direction of the extra atomic half-plane (hp) points along $\vec{p} = \vec{u} \times \vec{b}$ (also the slip-plane normal) where $\vec{u}$ is the dislocation line direction and $\vec{b}$ is the Burgers vector. For inserted half-planes to relieve the tensile epitaxial strain ($f = +0.64\%$), they must reside within the NiO film, requiring $\vec{p} \cdot \hat{n}_s > 0$. In bulk ionic rocksalt materials, plastic deformation occurs primarily on $\{110\}\langle110\rangle$ slip systems.\cite{haasenDislocationsPlasticityIonic1985} While $\{001\}$ is known to be a lower-energy subsidiary slip plane compared to $\{111\}$ in bulk crystals, it experiences zero resolved shear stress under (001) biaxial epitaxial strain. Consequently, to relieve epitaxial strain, we expect active slip planes to be $\{110\}$ first, followed by $\{111\}$. With a buried (001) epitaxial interface, glide of $\vec{b}=\frac{a}{2}\langle110\rangle$ TDs generates in-plane $\langle100\rangle$ and $\langle110\rangle$ MD segments gliding on $\{110\}$ and $\{111\}$ slip planes, respectively. Table~\ref{tab:s1} lists the slip plane normal $\vec{p}$, $\vec{b}$, and $\vec{u}$ expected when restricting strictly to tensile strain-relieving MDs of these slip families. 

\begin{table}[htbp]
\caption{Tensile-strain relieving misfit dislocations (MDs) considering two slip families in rocksalt epitaxial layers. The slip plane normal $(\vec{p})$, Burgers vector $(\vec{b})$, and MD line direction $(\vec{u})$ are listed for the case of a (001) interface. Calculated strain relief projections $(\vec{p} \cdot \hat{n}_s)$ and ECCI diffraction contrast values ($\vec{g} \cdot \vec{b}$ and $\vec{g} \cdot [\vec{b} \times \vec{u}]$) are listed for each MD, where $g_1=(200), g_2=(0\m20), g_3=(220),$ and $g_4=(2\m20)$. Column MD\# labels observable ECCI features with matching $|\vec{u}|$ and contrast sequence $\vec{g} \cdot [\vec{b} \times \vec{u}]$.}
\label{tab:s1}
\begin{ruledtabular}
\begin{tabular}{cccccccccccccc}
 & & & & & \multicolumn{4}{c}{$\vec{g} \cdot \vec{b}$} & \multicolumn{4}{c}{$\vec{g} \cdot [\vec{b} \times \vec{u}]$} & \\
\cline{6-9} \cline{10-13}
Family & $\vec{p}$ & $\vec{b}\ (a/2)$ & $\vec{u}$ & $\vec{p} \cdot \hat{n}_s$ & $g_1$ & $g_2$ & $g_3$ & $g_4$ & $g_1$ & $g_2$ & $g_3$ & $g_4$ & MD\# \\
\hline
\\
$\{111\}\langle 110\rangle$ & $[1\,1\,1]$ & $[1\,0\,\bar{1}]$ & $[1\,\bar{1}\,0]$ & 0.5 & 1 & 0 & 1 & 1 & -0.7 & 0.7 & -1.4 & 0 & 1a \\
 & $[1\,1\,1]$ & $[0\,1\,\bar{1}]$ & $[1\,\bar{1}\,0]$ & 0.5 & 0 & -1 & 1 & -1 & -0.7 & 0.7 & -1.4 & 0 & 1a' \\
 & $[\bar{1}\,\bar{1}\,1]$ & $[1\,0\,1]$ & $[1\,\bar{1}\,0]$ & 0.5 & 1 & 0 & 1 & 1 & 0.7 & -0.7 & 1.4 & 0 & 1b \\
 & [$\bar{1}\,\bar{1}\,1]$ & $[0\,1\,1]$ & $[1\,\bar{1}\,0]$ & 0.5 & 0 & -1 & 1 & -1 & 0.7 & -0.7 & 1.4 & 0 & 1b' \\
 & $[1\,\bar{1}\,1]$ & $[1\,0\,\bar{1}]$ & $[\bar{1}\,\bar{1}\,0]$ & 0.5 & -1 & 0 & -1 & -1 & -0.7 & -0.7 & 0 & -1.4 & 2a \\
 & $[1\,\bar{1}\,1]$ & $[0\,1\,1]$ & $[1\,1\,0]$ & 0.5 & 0 & -1 & 1 & -1 & -0.7 & -0.7 & 0 & -1.4 & 2a' \\
 & $[\bar{1}\,1\,1]$ & $[1\,0\,1]$ & $[\bar{1}\,\bar{1}\,0]$ & 0.5 & -1 & 0 & -1 & -1 & 0.7 & 0.7 & 0 & 1.4 & 2b \\
 & $[\bar{1}\,1\,1]$ & $[0\,1\,\bar{1}]$ & $[1\,1\,0]$ & 0.5 & 0 & -1 & 1 & -1 & 0.7 & 0.7 & 0 & 1.4 & 2b' \\
\\
\hline
\\
$\{110\}\langle 110\rangle$ & $[1\,0\,1]$ & $[1\,0\,\bar{1}]$ & $[0\,\bar{1}\,0]$ & 0.7 & 1 & 0 & 1 & 1 & -1 & 0 & -1 & -1 & 3a \\
 & $[\bar{1}\,0\,1]$ & $[1\,0\,1]$ & $[0\,\bar{1}\,0]$ & 0.7 & 1 & 0 & 1 & 1 & 1 & 0 & 1 & 1 & 3b \\
 & $[0\,1\,1]$ & $[0\,1\,\bar{1}]$ & $[1\,0\,0]$ & 0.7 & 0 & -1 & 1 & -1 & 0 & 1 & -1 & 1 & 4a \\
 & $[0\,\bar{1}\,1]$ & $[0\,1\,1]$ & $[1\,0\,0]$ & 0.7 & 0 & -1 & 1 & -1 & 0 & -1 & 1 & -1 & 4b \\
\end{tabular}
\end{ruledtabular}
\end{table}

\subsection{\label{sec:primary60} Primary $60^{\circ}$ Misfit Dislocations in Epitaxial NiO}

The strain relief characterized by $\vec{p} \cdot \hat{n}_s$ is larger for pure edge ($90^\circ$) MD segments of $\{110\}$ slip planes than mixed ($60^\circ$) MDs with $\{111\}$ slip planes. Although $\{110\}$ is the primary slip plane in bulk rocksalt, ECCI data show original MDs with $\vec{u}=\langle110\rangle$, consistent with $60^\circ$ MDs. Thus, the initial MD slip system in epitaxial NiO differs from bulk. Instead, the initial $\{111\}\langle110\rangle$ slip system acts as a kinetic pathway, eventually cross-slipping to form the higher strain-relieving pure edge MDs on $\{110\}$ planes at larger thicknesses. A similar observation is reported by Haidet et al.\ in epitaxial rocksalt PbSe under biaxial tension, where identical $60^\circ$ MDs are observed \cite{haidet_versatile_2023}. 

\begin{figure}[htbp]
\centering
\includegraphics[width=5.5in]{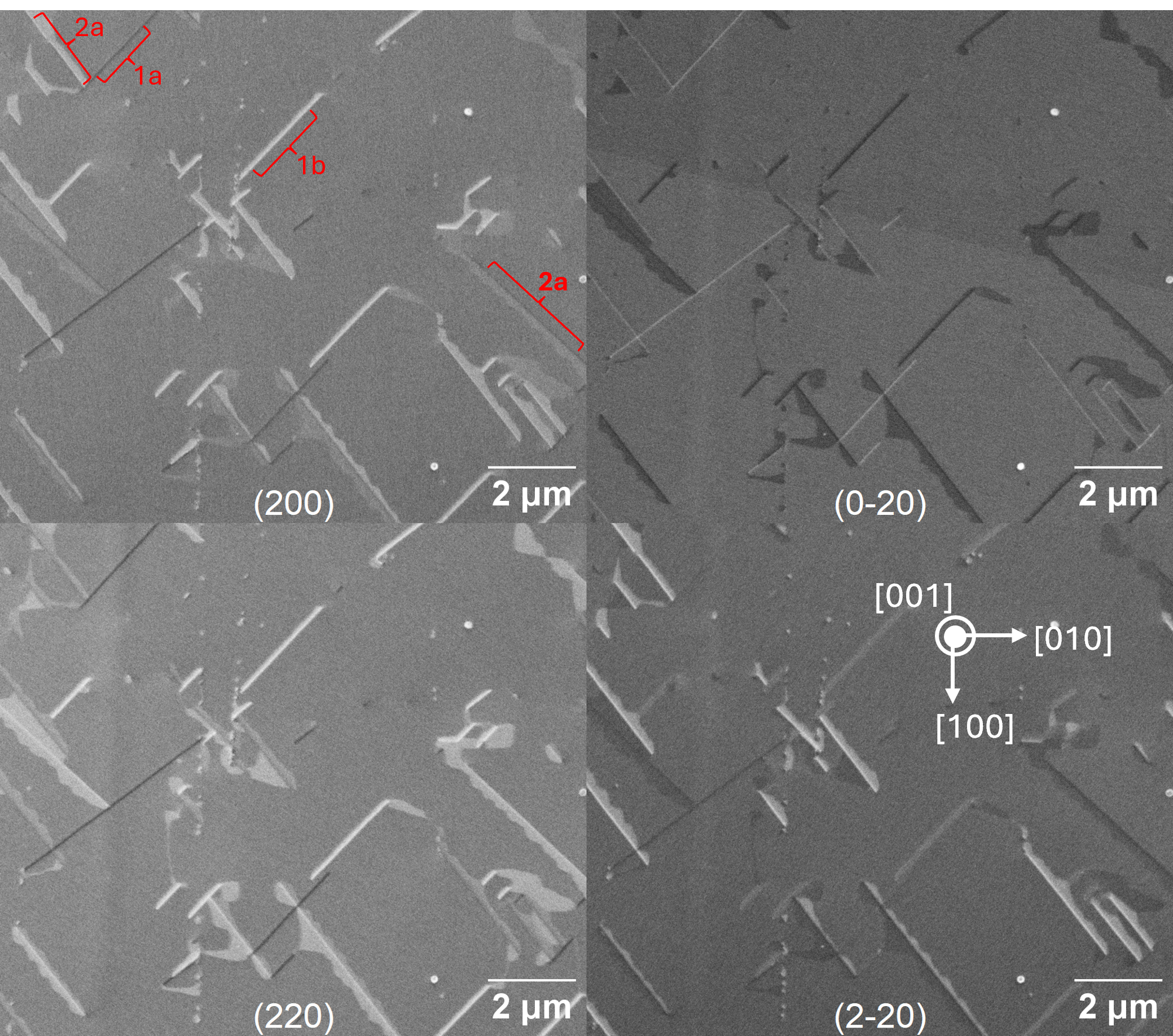}
\caption{ECCI images of 28~nm thick NiO (001) epilayers on MgO (001). Vector $\vec{g}$ and scale are listed in each panel. Red labels indicate MD line features distinguishable by line direction and ECCI contrast sequence (Table~\ref{tab:s1}).}
\label{ECCI28nm}
\end{figure}

ECCI data were collected from various sites on each sample at different $\vec{g}$. Figure~\ref{ECCI28nm} plots example ECCI images taken at distinct $\vec{g}$ for the 28~nm epilayer, showing almost exclusively $\vec{u}=\langle110\rangle$ consistent with $60^\circ$ MDs. The $\vec{u}=\langle110\rangle$ MDs divide into top-right/bottom-left MDs with $|\vec{u}|=\pm[1\bar{1}0]$ (MD1) and top-left/bottom-right features with $|\vec{u}|=\pm[110]$ (MD2). They exhibit symmetric (weak) extinction along $(220)$ and $(2\bar{2}0)$ (Fig.~\ref{ECCI28nm}), consistent with  $\vec{g} \cdot [\vec{b} \times \vec{u}] = 0$ invisibility criteria dominating for these MDs, rather than $\vec{g} \cdot \vec{b}=0$. The same is reported for SEM-ECCI of $60^\circ$ MDs in (001) zincblende structured III-V epilayers \cite{carnevaleRapidMisfitDislocation2014,boyer_correlation_2021}.

Each MD is subdivided further into two distinct ECCI variants per $|\vec{u}|$, changing contrast sign from dark in $(200)$ to light in $(0\bar{2}0)$ (labeled features a and b, respectively). This ECCI sequence matches relative contrast predicted by $\vec{g} \cdot (\vec{b} \times \vec{u})$ (Table~\ref{tab:s1}). Thus, four total ECCI-distinguishable MDs (MD 1a, MD 1b, MD 2a, MD 2b) are observed with $\vec{u}=\langle110\rangle$. Although there are eight possible $\vec{b}$, only four distinct ECCI variants are observable (two possible $\vec{b}$ assignments per ECCI feature). For example, MD 1a with $|\vec{u}|=[1\bar{1}0]$ exhibits ECCI sequence (dark, light, dark, extinct) for $g_{1-4}$ (Fig.~\ref{ECCI28nm}), matching $\vec{g} \cdot (\vec{b} \times \vec{u})$ for either $\vec{b}=[10\m1]$ or $[01\m1]$ (Table~\ref{tab:s1}). To differentiate these, we add a primed label, e.g MD 1a', for the second case of $\vec{b}$. These MDs form by surface half-loop nucleation, where the bottom segment of the expanding loop becomes an MD and outer segments become a pair of TDs (pure screw or $60^\circ$).\cite{freundThinFilmMaterials2004}

\subsection{\label{sec:cross-slip} Cross-Slip Formation of Pure Edge Misfit Dislocations}

As thickness increases, staircase MD traces appear with MD 1 and MD 2 segments connecting to long vertical or horizontal segments of $\vec{u}=\langle100\rangle$ (Fig.~\ref{ECCI-thickness}(d,e)). These MDs are explained by cross-slip of pure screw TDs, initially gliding on $\{111\}$ planes (tracing MDs with $\vec{u}=\langle110\rangle$) and switching slip planes to $\{110\}$ planes to form MDs with $\vec{u}=\langle100\rangle$. The crossover between dominant slip systems occurs between 30 and 35~nm (Fig.~\ref{ECCI-thickness}(k)). This illustrates that the initial $\{111\}\langle110\rangle$ slip system acts as a kinetically pathway to form the lower-energy (higher strain-relieving) $\{110\}\langle110\rangle$ $90^\circ$ MDs. This two-step process suggests an activation barrier for direct surface nucleation of half-loops in the $\{110\}$ family. Similar half-loop nucleation barriers are reported in other mismatched zinc-blende and rocksalt heteroepitaxial systems.\cite{legoues_measurement_1993, hull_misfit_1992, haidet_versatile_2023}

\begin{figure}[htbp]
\centering
\includegraphics[width=5.5in]{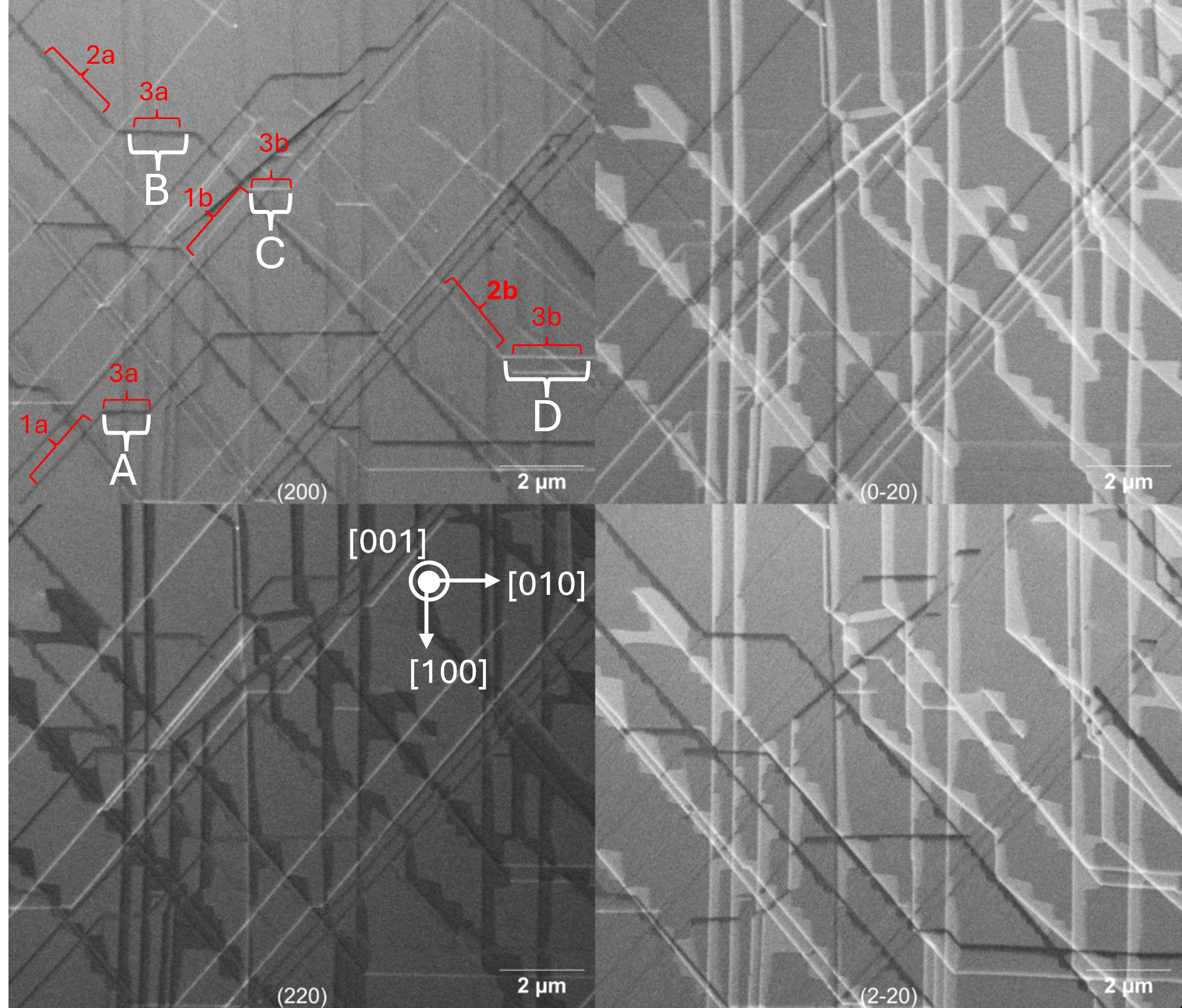}
\caption{ECCI images of 46~nm thick NiO (001) epilayers on MgO (001). Vector $\vec{g}$ and scale are listed in each panel. White labels (A--D) indicate examples of $|\vec{u}|=[0\m10]$ MDs with different cross-slip forming TDs, as indicated by connecting $|\vec{u}|=\langle110\rangle$ MDs. Red labels indicate $|\vec{u}|$ and ECCI contrast-distinguished MDs (Table~\ref{tab:s1}).}
\label{ECCI46nm}
\end{figure}

Representative ECCI images for the 46~nm epilayer showing long $\vec{u}=\langle100\rangle$ MD segments tracing from primary $\vec{u}=\langle110\rangle$ MDs are plotted in Fig.~\ref{ECCI46nm}. Right/left $|\vec{u}|=\pm[0\bar{1}0]$ and top/bottom $|\vec{u}|=\pm[100]$ MDs are labeled MD3 and MD4, respectively, exhibiting symmetric (strong) total extinction along $(200)$ and $(0\bar{2}0)$. Each line direction has two variants (dark or light) in $(200)$ or $(0\bar{2}0)$, labeled a and b. Thus, four distinct MD types (MD 3a, MD 3b, MD 4a, MD 4b) are observed with $\vec{u}=\langle100\rangle$. The strong invisibility of $\vec{u}=\langle100\rangle$ between $(200)$ and $(0\bar{2}0)$ is consistent with pure edge (Lomer) $90^\circ$ MDs, where simultaneously $\vec{g} \cdot \vec{b}=0$ and $\vec{g} \cdot (\vec{b} \times \vec{u})=0$ \cite{crimpDemonstration0Edge2001}. 

As shown in Table~\ref{tab:s1}, there are only four possible $\vec{b}$ for tensile-relieving $\{110\}\langle110\rangle$ MDs. ECCI contrast in this case uniquely identifies each $\vec{b}$ per $\vec{u}$. For example, MD3a with $\vec{u}=[0\bar{1}0]$ exhibits ECCI sequence (light, extinct, light, light) for $g_{1-4}$ (Fig.~\ref{ECCI46nm}), matching $\vec{g} \cdot (\vec{b} \times \vec{u})$ for strictly $\vec{b}=[101]$. Thus, the ECCI sequence for $\vec{u}=\langle100\rangle$ MDs provides unambiguous Burgers vector assignment.

\subsection{\label{sec:uniqueID} Unique Identification of Cross-Slip Pairs from $g$-Dependent ECCI}

Cross-slip of a gliding TD between $\{111\}$ and $\{110\}$ planes requires the TD to have the same Burgers vector $\vec{b}$ as that of the MD segments it forms. As seen in Table~\ref{tab:s1}, for each of the four MDs with $\vec{u}=\langle100\rangle$, there is a unique $\vec{b}$ with a unique ECCI contrast sequence. Because these MDs formed by TD cross-slip, the connecting MD traces with $\vec{u}=\langle110\rangle$ must share the same $\vec{b}$. Table~\ref{tab:s2} demonstrates this by sorting and matching MDs by $\vec{b}$ across slip families. For each of the four $\vec{b}$, there are two allowed cross-slip pairings, comprised of a $\vec{u}=\langle100\rangle$ MD (e.g. MD 3a), that can pair with one of two different $\vec{u}=\langle110\rangle$ that have the same $\vec{b}$, but different $\vec{u}$ and $\hat{n}$ (e.g. MD 1a or 2a). These are easily distinguished in ECCI by having orthogonal $\vec{u}$ (e.g. MD 1a $\perp$ MD 2a). This enables identification of not only the $\vec{b}$ of the cross-slipping TD, but also the particular slip planes involved. Under the constraint of cross-slip, we further can infer the $\vec{u}$ of the TD, since it must be  pure screw ($0^\circ$). Note that the cross-slip requirements dictate only that the gliding TD segment at the interface is purely screw-character; potential character changes in the remaining TD line closer to the free surface do not affect the interfacial MD formation mechanics.

\begin{table}[htbp]
\caption{Burgers vector $(\vec{b})$ matched MDs from Table~\ref{tab:s1} for tensile-strain relieving MDs in NiO films on MgO (001). MDs dinstinguishable by ECCI contrast and line direction are labeled MD\# for each slip family, see Table \ref{tab:s1}. Across rows, $\vec{b}$ are identical, defining allowed pairs of MDs formed by cross-slip of a pure screw TD.}
\label{tab:s2}
\begin{ruledtabular}
\begin{tabular}{c|ccccccccccccccc}
\multicolumn{1}{c}{} & \multicolumn{7}{c}{$\{110\}\langle 110\rangle$} & & \multicolumn{7}{c}{$\{111\}\langle 110\rangle$} \\
\cline{2-8} \cline{10-16}
\multicolumn{1}{c}{} \\
\multicolumn{1}{c}{$\vec{b}\ (a/2)$} & MD\# & $\vec{p}$ & $\vec{u}$ & $g_1$ & $g_2$ & $g_3$ & $g_4$ & & MD\# & $\vec{p}$ & $\vec{u}$ & $g_1$ & $g_2$ & $g_3$ & $g_4$ \\
\cline{1-1} \cline{2-8} \cline{10-16}
\\
\multirow{2}{*}{$[1\,0\,\bar{1}]$} & \multirow{2}{*}{3a} & \multirow{2}{*}{$[1\,0\,1]$} & \multirow{2}{*}{$[0\,\bar{1}\,0]$} & \multirow{2}{*}{-1} & \multirow{2}{*}{0} & \multirow{2}{*}{-1} & \multirow{2}{*}{-1} & & 1a & $[1\,1\,1]$ & $[1\,\bar{1}\,0]$ & -0.7 & 0.7 & -1.4 & 0 \\
 & & & &  & &  &  & & 2a & $[1\,\bar{1}\,1]$ & $[\bar{1}\,\bar{1}\,0]$ & -0.7 & -0.7 & 0 & -1.4 \\
\\
\hline
\\
\multirow{2}{*}{$[1\,0\,1]$} &\multirow{2}{*}{3b} & \multirow{2}{*}{$[\bar{1}\,0\,1]$} & \multirow{2}{*}{$[0\,\bar{1}\,0]$} & \multirow{2}{*}{1} & \multirow{2}{*}{0} & \multirow{2}{*}{1} & \multirow{2}{*}{1} & & 1b & $[\bar{1}\,\bar{1}\,1]$ & $[1\,\bar{1}\,0]$ & 0.7 & -0.7 & 1.4 & 0 \\
 & & & &  &  &  &  & & 2b & $[\bar{1}\,1\,1]$ & $[\bar{1}\,\bar{1}\,0]$ & 0.7 & 0.7 & 0 & 1.4 \\
\\
\hline
\\
\multirow{2}{*}{$[0\,1\,\bar{1}]$} & \multirow{2}{*}{4a} & \multirow{2}{*}{$[0\,1\,1]$} & \multirow{2}{*}{$[1\,0\,0]$} & \multirow{2}{*}{0} & \multirow{2}{*}{1} & \multirow{2}{*}{-1} & \multirow{2}{*}{1} & & 1a' & $[1\,1\,1]$ & $[1\,\bar{1}\,0]$ & -0.7 & 0.7 & -1.4 & 0 \\
 & & & &  &  &  &  & & 2b' & $[\bar{1}\,1\,1]$ & $[1\,1\,0]$ & 0.7 & 0.7 & 0 & 1.4 \\
\\
\hline
\\
\multirow{2}{*}{$[0\,1\,1]$} & \multirow{2}{*}{4b} & \multirow{2}{*}{$[0\,\bar{1}\,1]$} & \multirow{2}{*}{$[1\,0\,0]$} & \multirow{2}{*}{0} & \multirow{2}{*}{-1} & \multirow{2}{*}{1} & \multirow{2}{*}{-1} & & 1b' & $[\bar{1}\,\bar{1}\,1]$ & $[1\,\bar{1}\,0]$ & 0.7 & -0.7 & 1.4 & 0 \\
 & & & &  &  &  &  & & 2a' & $[1\,\bar{1}\,1]$ & $[1\,1\,0]$ & -0.7 & -0.7 & 0 & -1.4 \\
\end{tabular}
\end{ruledtabular}
\end{table}

To illustrate this, we identify four MD segments (A,B,C,D) with $|\vec{u}|=[0\m10]$ with cross-slip segments, in Fig. \ref{ECCI46nm}. A and B exhibit the same contrast sequence (MD 3a), although they have different cross-slip connecting MDs of either $|\vec{u}|=[1\m10]$ (MD 1a) or $[110]$ (MD 2a). Based on the ECCI contrast, sequence, and comparing to Table \ref{tab:s1}, A and B therefore have identical $\vec{b}=[10\m1]$ formed by glide on the $(101)$ plane. However, they differ only in the slip plane along which the TD cross-slipped. In the case of A, the $\vec{b}$ matching cross-slip MD segment has $|\vec{u}|=[1\m10]$, therefore of the $(111)$ slip plane, while for B, the $\vec{b}$ matching cross-slip MDs are $|\vec{u}|=[110]$, indicating $(\m11\m1)$. Thus, A forms a cross-slip pairing between features MD 3a-1a, while B pairs between MD 3a-2b. Similarly, for C and D (MD 3b), they exhibit an ECCI contrast sequence identifying $\vec{b}=[101]$ and a $(10\m1)$ plane. While their matching cross-slip MDs indicate a $(11\m1)$ slip plane for C (MD 3b-1b) and $(\m111)$ for D (MD 3b-2b).

Besides, the ability to identify the cross-slipping TD, the additional constraints enable unique identification of the $60^\circ$ MD burgers vectors, as implied by Table \ref{tab:s2}. More specifically, in Sec. \ref{sec:primary60}, we point out that for any particular $60^\circ$ MD, the observed line direction and ECCI contrast sequence cannot distinguish between two $\vec{b}$, e.g. MD 1a has two possible values of $\vec{b}=[10\m1]$ or $[01\m1]$, called MD 1a'. However, under the cross-slip constraint, the $\vec{b}$ is directly found from the unique $\vec{b}$ of the cross-slip pair. Specifically, in the case of MD 1a it can match with only MD 3a, and MD 1a' only with 4a, which are easily distinguishable from their observed $|\vec{u}|$.

In the discussion above we have not touched on the domain contrast, area features which spatially correlate with the dislocation regions. By superposition, ECCI shows a combination of this domain contrast with the MD contrast. This is expected to disrupt strict invisibility of MDs whenever a domain is superimposed on an MD in the SEM projected ECCI image.  In Figs. \ref{ECCI28nm} and \ref{ECCI46nm}  there are several examples of this. Careful examination of the images shows that the domain contrast is a milder perturbation to the contrast than the MDs, and is therefore, generally distinguishable. When MDs undergo extinction, remnant domain contrast can appear, but is differentiated by its different topology (uniform areas rather than sharp thin lines).

\section{Observation of Antiferromagnetic Domain Contrast in NiO Epitaxial Films on MgO (001)}
\label{sec:sec_3_afm_domain_contrast}

Broad, patchy domain features in plan-view ECCI disappear above the bulk Néel temperature of $T_N = 523\text{ K}$ (main text Fig.~1), unambiguously confirming their association with antiferromagnetic domain structure. Because ECCI measures in projection through the NiO epilayer, it is necessary to establish whether intensity variations originate from localized strain fields at finite-thickness domain walls or 3D volumetric domains.

Additionally, we must distinguish between twin (T) and spin (S) domains. The former are domains in which the NiO undergoes a rhombohedral contraction along a $\langle 111 \rangle$ axis normal to the $\{111\}$ easy-plane. This easy-plane contains ferromagnetically aligned \textit{intra-planar} spins, while \textit{inter-planar} spins are aligned antiferromagnetically. S-domains are a sub-variant, dividing a T-domain easy-plane further into regions with specific N\'eel vector easy-axes oriented along a $\langle 112 \rangle$ direction (in bulk NiO) within the $\{111\}$ plane of its T-domain. The combination of four unique T-domains ($T_k, k=1\dots4$) results in six unique T-domain walls (DWs). If we also consider S-domains, there are 12 unique combined TS magnetic domains yielding 66 unique TS-domain walls. Of these, 12 are pure S-DWs (with identical parent T-domains), leaving 54 mixed TS-DWs \cite{simonnot_atomic-scale_2026}. It should be noted that this S-domain structure is valid for bulk NiO, and is reported to change in thin films.\cite{schmittIdentifyingDomainwallSpin2023}

\subsection{Origin of ECCI Domain Contrast in NiO: $T$ vs $S$, and domain versus walls}

\begin{figure}[htbp]
\centering
\includegraphics[width=5.5in]{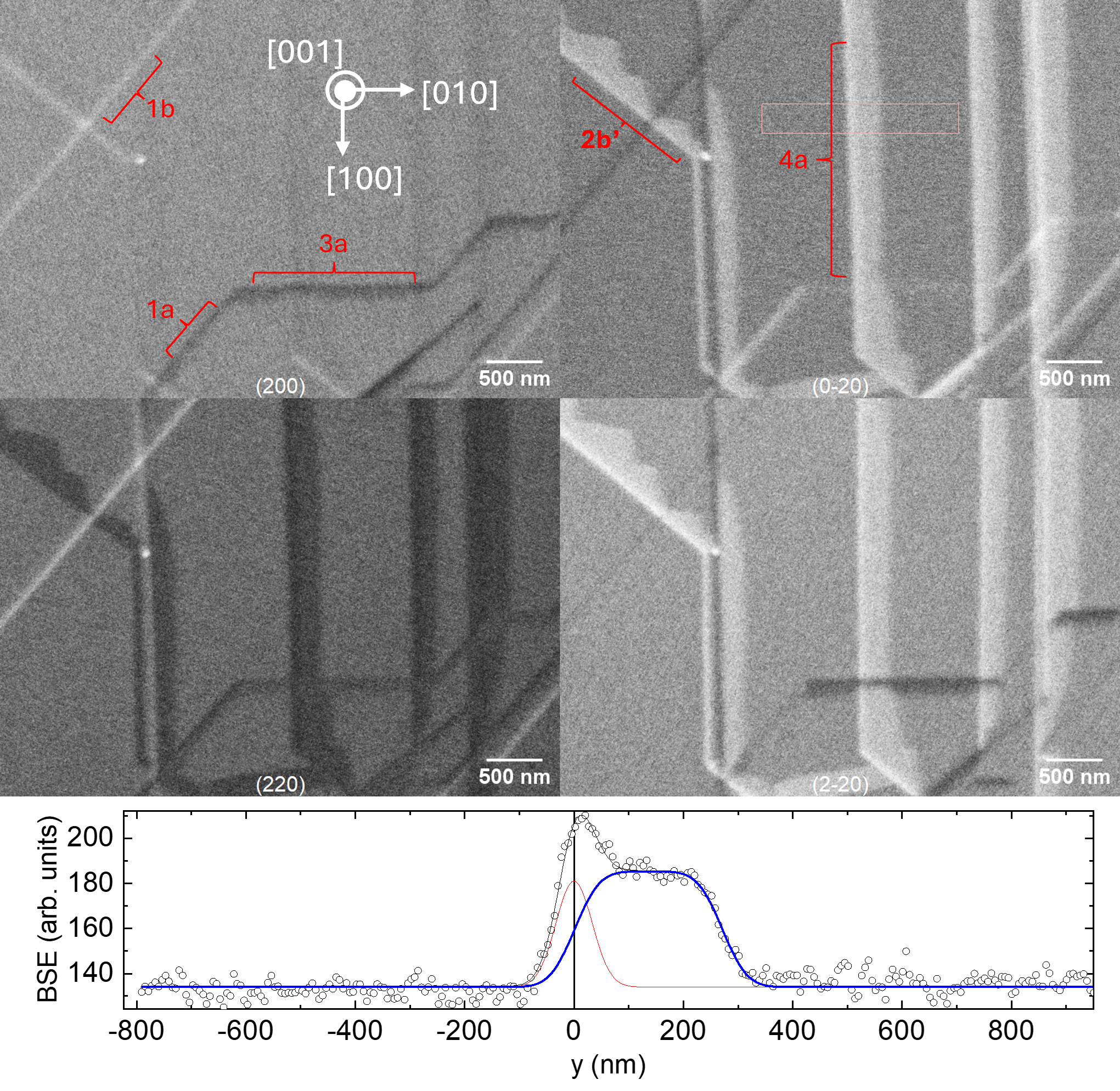}
\caption{ECCI images of 46~nm thick NiO (001) epilayers on MgO (001). Vector $\vec{g}$ and scale are listed in each panel. Red labels indicate $|\vec{u}|$ and ECCI contrast-distinguished MDs (Table~\ref{tab:s1}). T-domains are identified by first identifying the parent domain by noting the MD with no flanking domain, then referencing the appropriate ECCI contrast sequence in Table \ref{tab:s3}}.
\label{T-assignment}
\end{figure}

In bulk NiO and on free cleaved surfaces, competition between exchange energy and magnetoelastic anisotropy leads to extended $T$-domain walls, with literature values ranging from $\sim 150\text{ nm}$ up to $555\text{ nm}$ wide.\cite{weberMagnetostrictiveDomainWalls2003a, simonnot_atomic-scale_2026} If such wide transition zones were present in our films, smooth rotation of the easy-plane and lattice orientation across the wall would produce continuous intensity gradients in ECCI. However, experimental ECCI profiles across domain features (spanning $100\text{ nm}$ to $>400\text{ nm}$ laterally) exhibit flat, uniform intensity plateaus with sharp transition interfaces, see Fig. \ref{T-assignment}. A line cut taken across the y-axis of an MD 4a is plotted in the lower panel. The data are fitted to a Gaussian peak for the MD line contrast plus a Gaussian-convolved square pulse (using error function) for the domain feature. The fit results are shown as solid lines, yielding a FWHM of 33$\pm5$ nm for the MD line feature, and a T-domain edge width of 36$\pm3$ nm, values which may already be limited by the SEM resolution.  This indicates that $T$-domain walls in these epilayers are thinner than in bulk, which has previously been reported.\cite{schmittIdentifyingDomainwallSpin2023}.

Electron channeling sensitivity is governed by the rocking curve width $\Delta \theta_{\text{rocking}} \approx 1\text{--}5\text{ mrad}$ ($0.05^\circ\text{--}0.3^\circ$) \cite{picardTheoryDynamicalElectron2014}. Primary rhombohedral contraction ($\Delta\alpha \approx 4'$)\cite{rothNeutronOpticalStudies1960} generates a relative lattice tilt between $T$-domains, $\theta_T \approx 2.0\text{ to }3.5\text{ mrad}$. Because $\theta_T \sim \Delta \theta_{\text{rocking}}$, this structural tilt generates strong BSE contrast shifts. The secondary magnetostriction due to  spin-rotations within a given $T$-domain (\textit{easy-plane})   ($\lambda< 10^{-5}$)\cite{nakamichiAntiferromagneticMagnetostrictionCoO1961b}, is at least $5\times$ smaller than the rhombohedral distortion between different \textit{easy-planes}. The associated tilt is expected to be in the range, $\theta_S < 0.05\text{ mrad} \ll \Delta \theta_{\text{rocking}}$, indicating $S$-domains are invisible in ECCI. Thus, the contrast originates from 3D volumetric $T$-domains.

\subsection{Eigenstrain Tensors for Rhombohedral $T$-Domains}
\label{subsec:eigenstrain_tensors}

Above $T_N = 523\text{ K}$, NiO has a cubic rocksalt structure with $90^\circ$ inter-axis angles. Below $T_N$, rhombohedral contraction distorts pseudo-cubic cell angles to $\alpha_R = 90^\circ 4'$ ($\Delta\alpha \approx +1.16 \times 10^{-3}\text{ rad}$).\cite{slack_crystallography_1960}

The off-diagonal shear strain component $\varepsilon_{ij}^*$ ($i \neq j$) in the pseudo-cubic frame is given by $\varepsilon_{ij}^* = \varepsilon_s n_i n_j$, where $\hat{n}_k = \frac{1}{\sqrt{3}}(n_x, n_y, n_z)^T$ is the unit vector along the contraction axis and $\varepsilon_s < 0$ is spontaneous uniaxial strain. Defining the rhombohedral shear parameter $\eta \equiv \frac{1}{3}\varepsilon_s < 0$ ($\eta = -0.58 \times 10^{-3}$), the spontaneous transformation eigenstrain tensors $\boldsymbol{\varepsilon}^{*(k)}$ for the four $T$-domains consist strictly of off-diagonal shear components:
\begin{equation}
\label{eq:epsilon}
\boldsymbol{\varepsilon}^{*(1)} = \eta \left(\begin{smallmatrix} 0 & 1 & 1 \\ 1 & 0 & 1 \\ 1 & 1 & 0 \end{smallmatrix}\right), \quad
\boldsymbol{\varepsilon}^{*(2)} = \eta \left(\begin{smallmatrix} 0 & 1 & -1 \\ 1 & 0 & -1 \\ -1 & -1 & 0 \end{smallmatrix}\right), \quad
\boldsymbol{\varepsilon}^{*(3)} = \eta \left(\begin{smallmatrix} 0 & -1 & -1 \\ -1 & 0 & 1 \\ -1 & 1 & 0 \end{smallmatrix}\right), \quad
\boldsymbol{\varepsilon}^{*(4)} = \eta \left(\begin{smallmatrix} 0 & -1 & 1 \\ -1 & 0 & -1 \\ 1 & -1 & 0 \end{smallmatrix}\right)
\end{equation}

Note that background isotropic/biaxial strains ($\varepsilon_m \approx +0.64\%$, $\varepsilon_{zz} \approx -0.34\%$) are pre-existing elastic boundary conditions rather than phase transformation eigenstrains. Because normal strains are identical across all $T$-variants, they do not break fourfold crystallographic symmetry and drop out of Eshelby interaction energy calculations.

\subsection{NiO Parent $T$-Domain Nucleation and Defect-Induced Symmetry Breaking}
\label{subsec:parent_domain_nucleation}

Prior to MD formation, all four $\langle 111 \rangle$ contraction axes are symmetrically inclined relative to $(001)$ and energetically degenerate under biaxial strain. We therefore expect that upon cooling through $T_N$, thermal fluctuations will force the film to spontaneously select a single $\langle 111 \rangle$ contraction axis, establishing a single global matrix (parent) $T$-domain across coherent regions. In a ferromagnet, minimization of the dipolar field drives domain formation, but this field is absent in the antiferromagnet. The uniform stress field, before any defects form, provides no driving force for $T$-domain wall formation. However, secondary $T$-domains can form under a localized, asymmetric stress field if it is large enough to overcome the domain wall nucleation barrier. In partially relaxed films, localized stress fields ($\boldsymbol{\sigma}^{\text{MD}}$) surrounding MDs break local symmetry. The interaction between $\boldsymbol{\sigma}^{\text{MD}}$ and $\boldsymbol{\varepsilon}^*$, can therefore lower the local free energy on one flank of the dislocation, triggering nucleation of secondary (child) $T$-domains directly within the parent matrix. This MD and T-domain interaction will be described below using the Eshelby formalism. But first, we will explain how ECCI can detect T-domains.

\subsection{SEM-ECCI Contrast Modeling via Out-of-Plane Lattice Tilt}
\label{subsec:ecci_imaging_domains}

When a secondary domain variant $T_k$ is embedded within parent domain $T_p$, the difference in spontaneous strain tensors $\Delta \boldsymbol{\varepsilon}^{*(k,p)} = \boldsymbol{\varepsilon}^{*(k)} - \boldsymbol{\varepsilon}^{*(p)}$ breaks local symmetry. Constrained by the rigid $(001)$ interface, out-of-plane shear components $(\Delta \varepsilon_{xz}^{*(k,p)}, \Delta \varepsilon_{yz}^{*(k,p)})$ force a corrective backward rotation to maintain perfect registry, yielding the out-of-plane lattice tilt vector $\boldsymbol{\omega}^{(k,p)}$:
\begin{equation}
\boldsymbol{\omega}^{(k,p)} = \begin{pmatrix} -\Delta\varepsilon_{xz}^{*(k,p)} \\ -\Delta\varepsilon_{yz}^{*(k,p)} \\ 0 \end{pmatrix} = \begin{pmatrix} -(\varepsilon_{xz}^{*(k)} - \varepsilon_{xz}^{*(p)}) \\ -(\varepsilon_{yz}^{*(k)} - \varepsilon_{yz}^{*(p)}) \\ 0 \end{pmatrix}
\end{equation}

For an in-plane diffraction condition $\mathbf{g} = (h \, k \, 0)$, channeling contrast shift $C^{(k,p)}(\mathbf{g})$ relative to the parent matrix baseline is governed by:
\begin{equation}
C^{(k,p)}(\mathbf{g}) \propto \mathbf{g} \cdot \boldsymbol{\omega}^{(k,p)} = -(h \Delta\varepsilon_{xz}^{*(k,p)} + k \Delta\varepsilon_{yz}^{*(k,p)})
\end{equation}
where $\mathbf{g} \cdot \boldsymbol{\omega}^{(k,p)} > 0$ yields bright contrast, $< 0$ yields dark contrast, and $= 0$ corresponds to extinction. Table~\ref{tab:s3} lists tilt vectors and contrast projections for all 12 parent-child combinations across primary low-index in-plane diffraction vectors.

\begin{table}[htbp]
\centering
\caption{Relative out-of-plane lattice tilt vector $\boldsymbol{\omega}^{(k,p)}$ and ECCI contrast projection $\mathbf{g} \cdot \boldsymbol{\omega}^{(k,p)}$ for all 12 embedded secondary $T$-domain variants ($T_k$) relative to 4 parent matrix domains ($T_p$). Components and contrast values are expressed in normalized units of $|\eta| \approx 0.58 \times 10^{-3}$.}
\label{tab:s3}
\small
\begin{tabular}{ccccccc}
\hline\hline
\textbf{Parent Matrix} & \textbf{Secondary Variant} & \textbf{Lattice Tilt Vector} & \multicolumn{4}{c}{\textbf{Contrast Projection } $\mathbf{g} \cdot \boldsymbol{\omega}^{(k,p)}$} \\
\cline{4-7}
$T_p$ & $T_k$ & $\boldsymbol{\omega}^{(k,p)} / |\eta|$ & $\mathbf{g}_1 (200)$ & $\mathbf{g}_2 (0\bar{2}0)$ & $\mathbf{g}_3 (220)$ & $\mathbf{g}_4 (2\bar{2}0)$ \\
\hline
\multirow{3}{*}{$T_1 \; [111]$} 
 & $T_2 \; [11\bar{1}]$ & $(-2, -2, 0)^T$ & $-4$ & $+4$ & $-8$ & $0$ \\
 & $T_3 \; [\bar{1}11]$ & $(-2, \;\, 0, 0)^T$ & $-4$ & $0$ & $-4$ & $-4$ \\
 & $T_4 \; [1\bar{1}1]$ & $(\;\, 0, -2, 0)^T$ & $0$ & $+4$ & $-4$ & $+4$ \\
\hline
\multirow{3}{*}{$T_2 \; [11\bar{1}]$} 
 & $T_1 \; [111]$ & $(+2, +2, 0)^T$ & $+4$ & $-4$ & $+8$ & $0$ \\
 & $T_3 \; [\bar{1}11]$ & $(\;\, 0, +2, 0)^T$ & $0$ & $-4$ & $+4$ & $-4$ \\
 & $T_4 \; [1\bar{1}1]$ & $(+2, \;\, 0, 0)^T$ & $+4$ & $0$ & $+4$ & $+4$ \\
\hline
\multirow{3}{*}{$T_3 \; [\bar{1}11]$} 
 & $T_1 \; [111]$ & $(+2, \;\, 0, 0)^T$ & $+4$ & $0$ & $+4$ & $+4$ \\
 & $T_2 \; [11\bar{1}]$ & $(\;\, 0, -2, 0)^T$ & $0$ & $+4$ & $-4$ & $+4$ \\
 & $T_4 \; [1\bar{1}1]$ & $(+2, -2, 0)^T$ & $+4$ & $+4$ & $0$ & $+8$ \\
\hline
\multirow{3}{*}{$T_4 \; [1\bar{1}1]$} 
 & $T_1 \; [111]$ & $(\;\, 0, +2, 0)^T$ & $0$ & $-4$ & $+4$ & $-4$ \\
 & $T_2 \; [11\bar{1}]$ & $(-2, \;\, 0, 0)^T$ & $-4$ & $0$ & $-4$ & $-4$ \\
 & $T_3 \; [\bar{1}11]$ & $(-2, +2, 0)^T$ & $-4$ & $-4$ & $0$ & $-8$ \\
\hline\hline
\end{tabular}
\end{table}

\section{Eshelby-based Magnetoelastic Energy Model for Dislocation-Induced Antiferromagnetic $T$-Domains}
\label{sec:eshelby_model}

Antiferromagnetic $T$-domains in NiO form with reproducible and complex patterns near specific misfit dislocation (MD) features (main text Figs.~1--4). $T$-domains are rarely observed in the thinest films containing few MDs (Fig.~\ref{ECCI-thickness}(a)), but increase in area fraction with film thickness in direct correspondence with MD density. In many cases, $T$-domains appear as rectangular or sawtooth features immediately adjacent to MDs, extending across their line length at consistent positions along their flanks. Given that $T$-domains are defined by spontaneous rhombohedral eigenstrains (Sec.~\ref{sec:sec_3_afm_domain_contrast}), their spatial correlation with MDs is surely driven by the local dislocation stress fields.

In this section, we define the expected stress tensors due to the observed $\{111\}\langle110\rangle$ and $\{110\}\langle110\rangle$ MDs and apply the Eshelby formalism to calculate the magnetoelastic interaction energy change for T-domains forming in the NiO in the vicinity of the high-stress dislocation cores. The modeled energy landscape shows a strongly preferred T-domain forming on one flank of every 60$^{\circ}$ MD. The relative strength of the MD/T-domain interaction is compared with other magnetic order related energies, and is found to be at least one order of magnitude larger than the domain wall energy penalty, indicating the shear stresses near the MD core can easily dominate the antiferromagnetic domain structure. The predicted T-domain MD formation is compared to $\vec{g}$-dependent ECCI images, enabling identification of the parent T-domain and all observed flanking domains.

\subsection{Stress Field of Misfit Dislocations}
\label{subsec:md_stress_fields}

The stress field of MDs is defined in the natural Volterra coordinate frame $(x_V, y_V, z_V)$, at the dislocation line such that $\hat{x}_V$ aligns with the line direction $\vec{u}$. The physical Burgers vector $\vec{b}$ is decomposed into screw and edge components. We define $\hat{y}_V$ to lie strictly along the edge component of the Burgers vector, and $\hat{z}_V = \hat{x}_V \times \hat{y}_V$. This rotated frame forces $b_z=0$ , i.e. $\vec{b}_V = (b_x, b_y, 0)$.

For a straight dislocation in an isotropic, infinite medium with shear modulus $\mu$ and Poisson's ratio $\nu$, the local Volterra stress tensor $\boldsymbol{\sigma}^V(y_V, z_V)$ is given by the standard superposition of pure screw and pure edge fields:\cite{andersonTheoryDislocations2017}
\begin{align}
\label{eq:sigma_volterra}
\sigma_{xx}^V &= -\frac{\mu \nu}{\pi(1-\nu)} \frac{b_y z_V}{y_V^2 + z_V^2}, \notag \\
\sigma_{yy}^V &= -\frac{\mu}{2\pi(1-\nu)} \frac{b_y z_V (3y_V^2 + z_V^2)}{(y_V^2 + z_V^2)^2}, \notag \\
\sigma_{zz}^V &= \frac{\mu}{2\pi(1-\nu)} \frac{b_y z_V (y_V^2 - z_V^2)}{(y_V^2 + z_V^2)^2}, \\
\sigma_{yz}^V &= \frac{\mu}{2\pi(1-\nu)} \frac{b_y y_V (y_V^2 - z_V^2)}{(y_V^2 + z_V^2)^2}, \notag \\
\sigma_{xy}^V &= -\frac{\mu b_x}{2\pi} \frac{z_V}{y_V^2 + z_V^2}, \notag \\
\sigma_{xz}^V &= \frac{\mu b_x}{2\pi} \frac{y_V}{y_V^2 + z_V^2}. \notag
\end{align}
Before using this for any calculations, $\boldsymbol{\sigma}^V(y_V, z_V)$ is rotated into the crystal coordinate frame with tensor transformation: $\boldsymbol{\sigma}^{MD} = \mathbf{R} \boldsymbol{\sigma}^V \mathbf{R}^T$, where $\mathbf{R} = [\hat{x}_V \mid \hat{y}_V \mid \hat{z}_V]$ is constructed from the basis vectors of the Volterra frame expressed as crystal coordinates.\cite{de_graef_structure_2012} 

The above ignores the heterointerface with elastic modulus mismatch between the thin film and substrate, as well as the nearby free surface which must relax out-of-plane stresses.\cite{freundThinFilmMaterials2004} The elastic mismatch parameters ($\alpha \approx -0.16, \beta \approx -0.04$ for $\text{NiO}/\text{MgO}$) introduce minor stress corrections. The infinite-medium approximation remains appropriate here, as $T$-domain nucleation is dominated by the strong $1/r$ dependence of the primary Volterra core field.

\subsection{Eshelby Magnetoelastic Interaction Energy and Flank-Specific $T$-Domain Stabilization}
\label{subsec:eshelby_energetics}

\begin{figure}[htbp]
\centering
\includegraphics[width=6.5in]{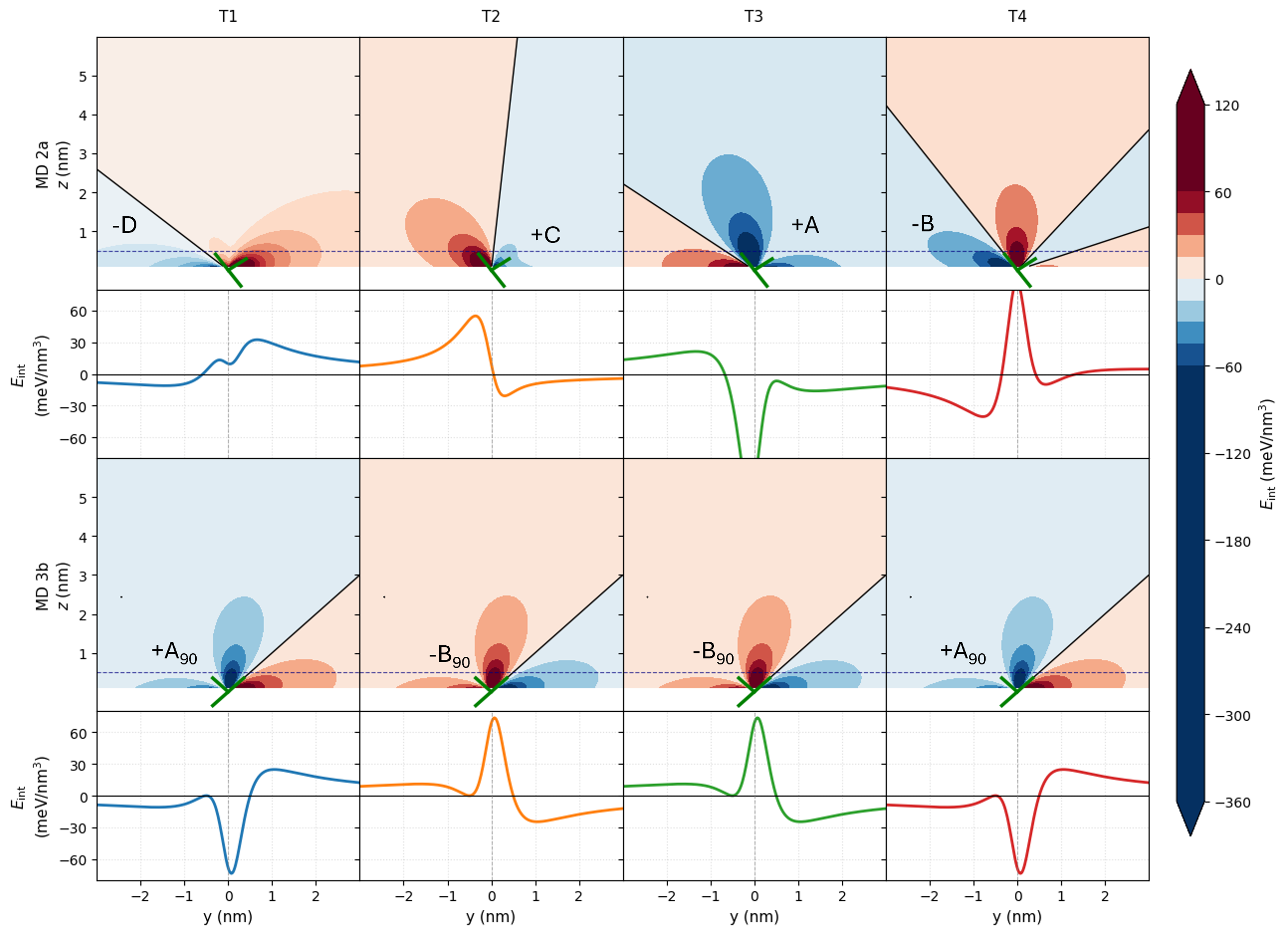}
\caption{Magneto-Elastic interaction ($E_{int}$) between misfit dislocations (MDs) and antiferromagnetic T-domains in NiO(001) on MgO calculated by Eshelby formalism, Eq. \ref{eq:eshelby}. Color maps plot $E_{int}(y,z)$ around the different MDs (top and bottom row) for the four T-domains(across columns). MD line vector $(\vec{u})$ is parallel to $x$, out of page, film/substrate interface is at $z=0$, $\perp$ symbol (green) gives the direction of the half-plane ($\vec{p}$) and slip-plane, black lines divide between $E_{int}<0$ (blue) and $E_{int}>0$ (red) regions, and blue dashed line indicates $z$ position of 1D slice, plotted below each panel. Case A shows the most favorable T-domain for MD 2a, where $\hat{x}\parallel[\bar{1}\bar{1}0]$ which corresponds to the $T_3$ $(\bar{1}11)$ easy-plane being most nearly aligned with the extra half-plane of the MD $(\bar{1}12)$, making it likely to nucleate on the $+y$ flank of MD 2a. For the pure edge MD 3b, $\hat{x}\parallel[0\bar{1}0]$, and both $T_1$ and $T_4$ are equally likely to form on the $+y$ flank.}
\label{fig:S6}
\end{figure}

We employ an Eshelby magnetoelastic framework, where the  MD stress field $\boldsymbol{\sigma}^{\text{MD}}(y,z)$ (Sec. \ref{subsec:md_stress_fields}) does mechanical work on the crystal during the spontaneous/stress-free transformation (cubic to rhombohedral) characterized by the T-domain eigenstrain tensor $\boldsymbol{\varepsilon}^*$ (Eq. \ref{eq:epsilon}). This adds a contribution to the free energy that reflects the magnetoelastic interaction of MDs with T-domains, $E_{int}$, given by the Eshelby double-dot product:
\begin{equation}
E_{int}(y,z) = -\boldsymbol{\sigma}^{\text{MD}}(y,z) : \boldsymbol{\varepsilon}^* = -\sum_{i,j} \sigma_{ij}^{\text{MD}}(y,z) \, \varepsilon_{ij}^*
\end{equation}
Thee diagonal components of $\boldsymbol{\varepsilon}^*$ are zero, such that normal stresses ($\sigma_{xx}, \sigma_{yy}, \sigma_{zz}$) drop out of $E_{int}(y,z)$. This simplifies the magnetoelastic interaction to include only shear stress components:
\begin{equation}
E_{int}^{(k)}(y,z) = -2 \left[ \sigma_{xy}^{\text{MD}}(y,z) \, \varepsilon_{xy}^{*(k)} + \sigma_{yz}^{\text{MD}}(y,z) \, \varepsilon_{yz}^{*(k)} + \sigma_{xz}^{\text{MD}}(y,z) \, \varepsilon_{xz}^{*(k)} \right]
\label{eq:eshelby}
\end{equation}
The strong stress field near the MD core leads to a substantial interaction energy with the T-domains. This is illustrated by plotting Eq. \ref{eq:eshelby} in the immediate vicinity of the MD core, Fig. \ref{fig:S6}, in a local MD viewing frame (not to be confused with the Volterra frame), where $\hat{x}\parallel\vec{u}$, $\hat{z}\parallel[001]$, and $\hat{y}=\hat{z}\times\hat{x}$. The plot is produced using a lattice parameter $a = 0.417$ nm, $\mu=100.0$ GPa, $\nu=0.3$, and $\eta=-.00058$. All 48 cases (4 T-domains and 12 MDs) were examined, with representative cases shown in Fig. \ref{eq:eshelby} for a 60$^{\circ}$ (MD 1b) and a 90$^{\circ}$ (MD 3b). 

For all of the 60$^{\circ}$ MDs, each of the T-domains produces a different $E_{int}$, where one of the T-domains shows deep energy minima just at the core, labeled case $\pm$A, where the sign indicates the most stable $y$-flank. In the highly favorable case A, the T-domain planes are closest aligned (19.47$^{\circ}$) to the half-plane (hp) of the MD. More specifically,  $\hat{n}_k\cdot\vec{p}$ is largest, where $\hat{n}_k$ is the $T_k$ domain plane normal (the rhombohedral contraction axis). Essentially, the hp wedge is best aligned with one particular T-domain. A second special case, labeled B, occurs when the plane of the T-domain is exactly equal to othe slip plane of the MD. In that case the $\{111\}$ planes of the T-domain are orthogonal to othe hp. This results in a strong change in $E_{int}$, where a massive energy peak occurs just above the MD core near the hp. Yet, below the slip plane, a strong energy minimum occurs just to one side of the MD. The two other T-domains show a much lower interaction energy with the MD, as these cases (C and D) are T-domains that are not parallel to the MD line. However, they still display strong interaction energies (both positive and negative) asymmetrically across the MD flank. 

For the pure edge MDs, there are only 2 distinct cases, labeled A$_{90}$ and B$_{90}$. This is because for the MDs of the $\{101\}\langle101\rangle$ slip system, there are always 2 T-domains with the same orientation with respect to the hp, yielding only 2 distinct cases. Notably, the interaction energies are lower for these MDs due to the poorer alignment of the hp with the T-domain $\{111\}$ planes than in the case of the 60$^{\circ}$ MDs.

The 90$^{\circ}$ MDs show just 2 unique $E_{int}$ cases ($\pm$A90, and $\pm$B90), since there are 2  pairs of T-domains with identical orientation to $\vec{u}$ and $\vec{b}$. and produce identical $E_{int}$, differing only in the sign for each T-domain. The 90$^{\circ}$ MDs therefore do not fully break the T-domain degeneracy. Notably, the interaction energies are lower for these MDs due to the poorer alignment of the hp with the T-domain $\{111\}$ planes than in the case of the 60$^{\circ}$ MDs.

\subsection{Relative Strength of Dislocation/T-domain Interaction}
\label{subsec:relative-strength}

To determine whether this magnetoelastic coupling is sufficient to nucleate secondary $T$-domains at MD flanks, we evaluate the competing energy scales governing NiO domain formation:
\begin{enumerate}
    \item \textbf{Magnetoelastic Core Interaction Energy Density ($E_{\text{int}}$):} Shear stresses surrounding the MD core ($r \approx 1\text{--}5\text{ nm}$) generate massive local energy density changes. As mapped in Fig.~\ref{fig:S6}, a $60^\circ$ MD in case A, where the T-domain is most closely aligned with the extra-half-plane, leads to an $E_{\text{int}}$ dropping below $-60\text{ meV/nm}^3$ reaching a minimum of $-120\text{ meV/nm}^3$ just above the core. In case B, where the $T$-domain matches the slip plane, there is a large maximum of $E_{\text{int}}$ just above the core reaching nearly $+100\text{ meV/nm}^3$, while immediately to one $y$ side of the core, there is a $-30\text{ meV/nm}^3$ minimum.
    \item \textbf{Magnetocrystalline Anisotropy ($K_1$):} In-plane magnetocrystalline anisotropy in NiO is relatively weak, $K_1 \approx 10^3\text{--}10^4\text{ J/m}^3$ ($10^{-3}\text{--}10^{-2}\text{ meV/nm}^3$) \cite{slack_crystallography_1960, weberMagnetostrictiveDomainWalls2003a}. 
    \item \textbf{Domain Wall Nucleation Barrier ($\gamma_{\text{DW}}$):} Forming a rhombohedral $T$-domain wall incurs an interface energy penalty of $\gamma_{\text{DW}} \approx 1\text{--}3\text{ mJ/m}^2$ \cite{weberMagnetostrictiveDomainWalls2003a, simonnot_atomic-scale_2026}. For a secondary domain slab extending over lateral width $w \approx 10\text{--}30\text{ nm}$, the volumetric domain wall energy penalty is $\gamma_{\text{DW}}/w \approx (0.3\text{--}1.0) \times 10^5\text{ J/m}^3$ ($0.02\text{--}0.06\text{ meV/nm}^3$). 
\end{enumerate}

Because the local Eshelby energy reduction near the core ($|E_{\text{int}}| \sim 100\text{--}120\text{ meV/nm}^3$) exceeds the volumetric domain wall penalty ($\gamma_{\text{DW}}/w \sim 0.02\text{--}0.06\text{ meV/nm}^3$) by over three orders of magnitude, local dislocation shear fields supply an overwhelming thermodynamic driving force that easily overcomes the $T$-domain wall nucleation barrier. 

Note on continuum limits: While our continuum linear elastic model calculates peak interaction energies of $-120$~meV/nm$^3$ within 1 nm of the core, continuum elasticity breaks down at these discrete atomistic scales where core chemistry dominates. However, even at a distance of 3 nm from the core, the purely magnetoelastic driving force remains $-\sim 10$~meV/nm$^3$, which still vastly exceeds the domain wall penalties ($\sim 0.05$~meV/nm$^3$). Therefore, while exact core atomistics require future first-principles simulations, the long-range elastic strain overwhelmingly dictates the local thermodynamics.

\subsection{Unique T-domain assignment from observed dislocation flanking and g-dependent ECCI}
\label{subsec:T-assignment}

A complete theoretical treatment explaining how this MD core interaction results in the observed T-domain formation should include a full phase field model, but this is beyond the scope of this Supplementary. Instead, here we discuss a qualitative prediction of $E_{int}$ on the stabilization of variant T-domains near/at MDs, and discuss the possible nucleation and growth mechanism of T-domains in NiO epilayers.

\begin{figure}[htbp]
\centering
\includegraphics[width=4.5in]{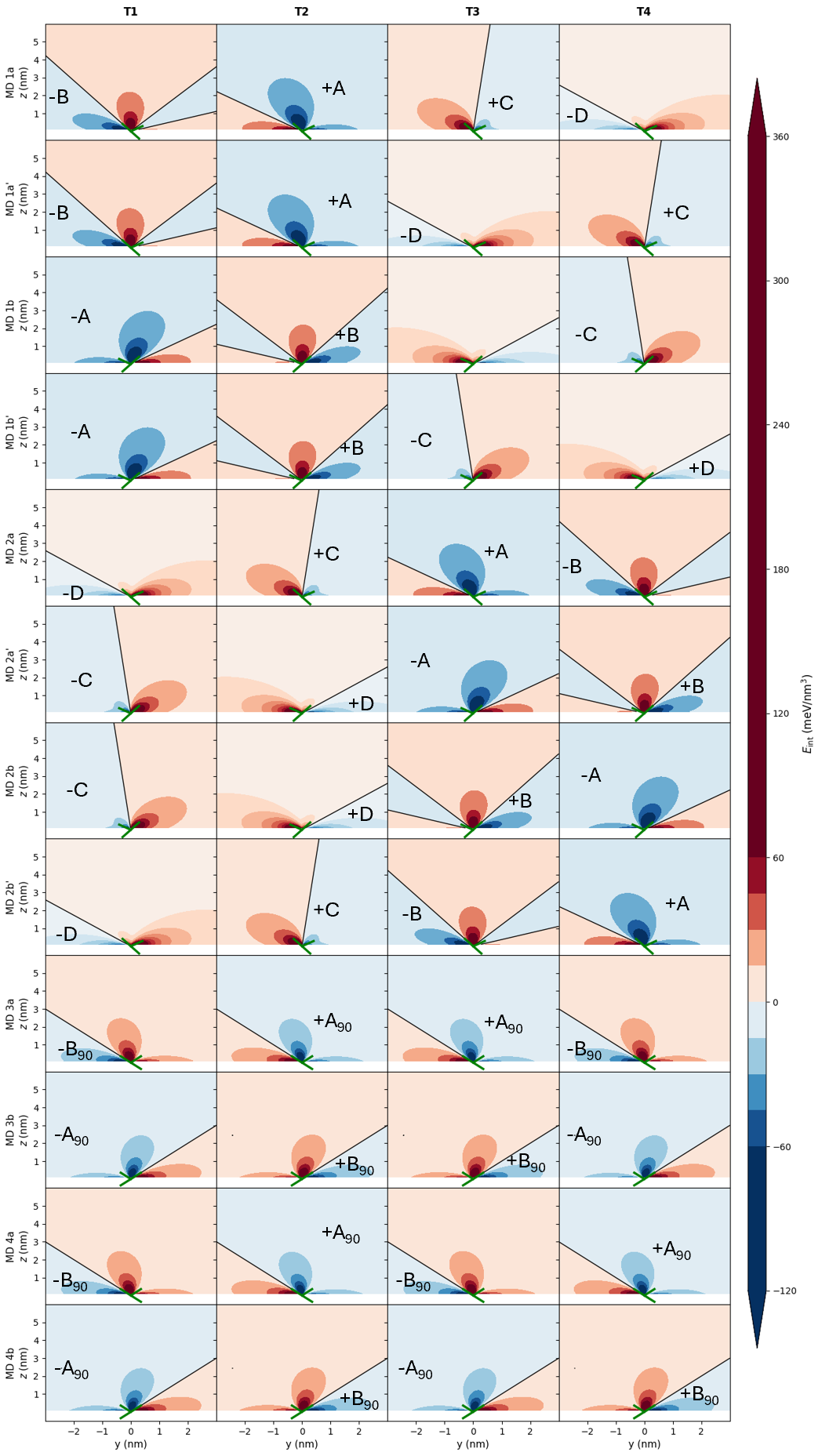}
\caption{All 48 cases of MDs flanked by T-domains in NiO/MgO (001)}
\label{fig:S7}
\end{figure}

First we review the  48 possible $E_{int}$  cases in our films (12 MDs and 4 T-domains) , plotted in Fig. \ref{fig:S7}. Each panel is labeled one of the 12 unique MD/T-domain cases: $\pm$A,$\pm$B,$\pm$C,$\pm$D, $\pm$A$_{90}$, and $\pm$B$_{90}$, as described in Sec. \ref{subsec:eshelby_energetics} and Fig. \ref{fig:S6}. The results are compiled into Table \ref{tab:s4}. Clearly case A will dominate T-domain formation at MD cores. In the SEM-ECCI data we observe mostly MDs flanking one side of an MD, yet in some cases different domains are observed across the MD line. For those, it is worth considering the secondary preferred T-domain after case A. We note that the negative $E_{int}$ region for case B complements that of case A, such that a T-domain of case A bounded by a T-domain of case B is possible. The other cases, C and D could also act as secondary domains, however, their $E_{int}$ reductions are not as large as that of B. With these considerations, we tabulate case A (in bold) as the predicted MD flanking T-domain forming on the $\pm y$ flank for all 12 possible MDs. The secondary case B (in grey), is also included to predict likely double-flanking domain cases.
\begin{table}[htbp]
\centering
\caption{Magnetoelastic energy ($E_{\text{int}}$) cases, predicted MD flanking $T$-domains, and observed ECCI domain flanks at misfit dislocations.}
\label{tab:s4}
\setlength{\tabcolsep}{7pt}
\begin{tabular}{lcccccccc}
\hline\hline
 & \multicolumn{4}{c}{$T$-domain} & \multicolumn{4}{c}{MD flanking $T$-domain} \\
\cline{2-5} \cline{6-9}
 & & & & & \multicolumn{2}{c}{Prediction} & \multicolumn{2}{c}{ECCI observed} \\
\cline{6-7} \cline{8-9}
MD\# & $T_1$ & $T_2$ & $T_3$ & $T_4$ & A & B & 26\,nm & 46\,nm \\
\hline
1a   & $-\text{B}$ & $+\text{A}$ & $+\text{C}$ & $-\text{D}$ & $\mathbf{+T_2}$ & \textcolor{gray}{$-T_1$} & $0$ & \\
1a$'$  & $-\text{B}$ & $+\text{A}$ & $-\text{D}$ & $+\text{C}$ & $\mathbf{+T_2}$ & \textcolor{gray}{$-T_1$} & $0$ & \\
1b   & $-\text{A}$ & $+\text{B}$ & $+\text{D}$ & $-\text{C}$ & $\mathbf{-T_1}$ & \textcolor{gray}{$+T_2$} & $-d_1$ & $0$ \\
1b$'$  & $-\text{A}$ & $+\text{B}$ & $-\text{C}$ & $+\text{D}$ & $\mathbf{-T_1}$ & \textcolor{gray}{$+T_2$} & & \\
2a   & $-\text{D}$ & $+\text{C}$ & $+\text{A}$ & $-\text{B}$ & $\mathbf{+T_3}$ & \textcolor{gray}{$-T_4$} & $+d_3\,(-d_2)$ & $+d_4$ \\
2a$'$  & $-\text{C}$ & $+\text{D}$ & $-\text{A}$ & $+\text{B}$ & $\mathbf{-T_3}$ & \textcolor{gray}{$+T_4$} & $-d_3\,(+d_2)$ & \\
2b   & $-\text{C}$ & $+\text{D}$ & $+\text{B}$ & $-\text{A}$ & $\mathbf{-T_4}$ & \textcolor{gray}{$-T_3$} & $-d_2$ & $-d_5$ \\
2b$'$  & $-\text{D}$ & $+\text{C}$ & $-\text{B}$ & $+\text{A}$ & $\mathbf{+T_4}$ & \textcolor{gray}{$-T_3$} & $+d_2$ & $+d_5$ \\
\hline
3a   & $-\text{B}_{90}$ & $+\text{A}_{90}$ & $+\text{A}_{90}$ & $-\text{B}_{90}$ & $\mathbf{+T_2\text{ or }T_3}$ & \textcolor{gray}{$-T_1\text{ or }T_4$} & & $+d_1$ \\
3b   & $-\text{A}_{90}$ & $+\text{B}_{90}$ & $+\text{B}_{90}$ & $-\text{A}_{90}$ & $\mathbf{-T_1\text{ or }T_4}$ & \textcolor{gray}{$-T_2\text{ or }T_3$} & $-d_1$ & $0$ \\
4a   & $-\text{B}_{90}$ & $+\text{A}_{90}$ & $-\text{B}_{90}$ & $+\text{A}_{90}$ & $\mathbf{+T_2\text{ or }T_4}$ & \textcolor{gray}{$-T_1\text{ or }T_3$} & & $+d_2$ \\
4b   & $-\text{A}_{90}$ & $+\text{B}_{90}$ & $-\text{A}_{90}$ & $+\text{B}_{90}$ & $\mathbf{-T_1\text{ or }T_3}$ & \textcolor{gray}{$+T_2\text{ or }T_4$} & & \\
\hline\hline
\end{tabular}

\vspace{0.4cm}

\setlength{\tabcolsep}{10pt}
\begin{tabular}{ccccc}
\hline\hline
\multicolumn{5}{c}{\textbf{ECCI domains observed}} \\
\hline
$d$\# & $g_1$ & $g_2$ & $g_3$ & $g_4$ \\
\hline
$d_1$ & $+$ & $-$ & $+$ & $0$ \\
$d_2$ & $+$ & $0$ & $+$ & $+$ \\
$d_3$ & $0$ & $-$ & $+$ & $-$ \\
$d_4$ & $-$ & $0$ & $-$ & $-$ \\
$d_5$ & $0$ & $+$ & $-$ & $+$ \\
\hline\hline
\end{tabular}
\end{table}

Finally, we compare the prediction to ECCI for the 28nm and 46nm sample data shown in Figs. \ref{ECCI28nm} and \ref{ECCI46nm}. In both films we find full agreement between the predicted and observed flanking direction and T-domain, as identified uniquely by their $\vec{g}$-dependent contrast sequences. The observed domains are listed in the sub-panel of Table \ref{tab:s4} as $d_{1-5}$ with the their observed contrast at each $\vec{g}$ condition. The first three ($d_{1-3}$) are observed in the 28nm sample, while two types types are observed in the 46nm film, $d_{4,5}$. 

First, we note the specific $y$-flank and ECCI contrast for domains occurring at MDs. These observations are listed in main Table \ref{tab:s4}. In the 28nm sample, for example, $-d_1$ indicates that this ECCI distinguishable domain is observed to flank the $-y$ side of MD 1b. The $E_{int}$ prediction shows that case A indeed occurs on the $-y$ flank, and is expected to be $T_1$ for this dislocatioon. To verify that assignment, we compare the observed contrast sequence $g_{1-4}$ for $d_1$ (+,-,+,0) with the predicted contrast sequences for specific $T_k$ domains, listed in Table \ref{tab:s3}. The observed $d_1$ contrast sequence uniquely matches the sequence for $T_1$ when $T_p=T_2$. Using this method, we identify $d_2$ uniquely as $T_4$ when $T_p=T_2$, confirming the initial parent domain assignment. Finally, for the 28 nm sample $d_3$ is identified as $T_3$, when $T_p=T_2$ (although for this particular sequence, its ECCI contrast alone, does not uniquely identify the parent domain). Lastly, we note that for the 28nm sample, ECCI images of MD 1a or 1a' show no flanking domains, i.e. the  0 listed in Table \ref{tab:s4}. Furthermore MD 1a or 1a' are the only dislocations observed in this sample with no flanking domains. According to the $E_{int}$ prediction, these MDs should stabilize $T_2$. This indicates $T_2=T_p$, as the T-domain ECCI contrast is a relative tilt from the parent domain, which is 0 by definition for the parent domain, see Sec. \ref{subsec:ecci_imaging_domains}. 

\begin{figure}[htbp]
\centering
\includegraphics[width=6.5in]{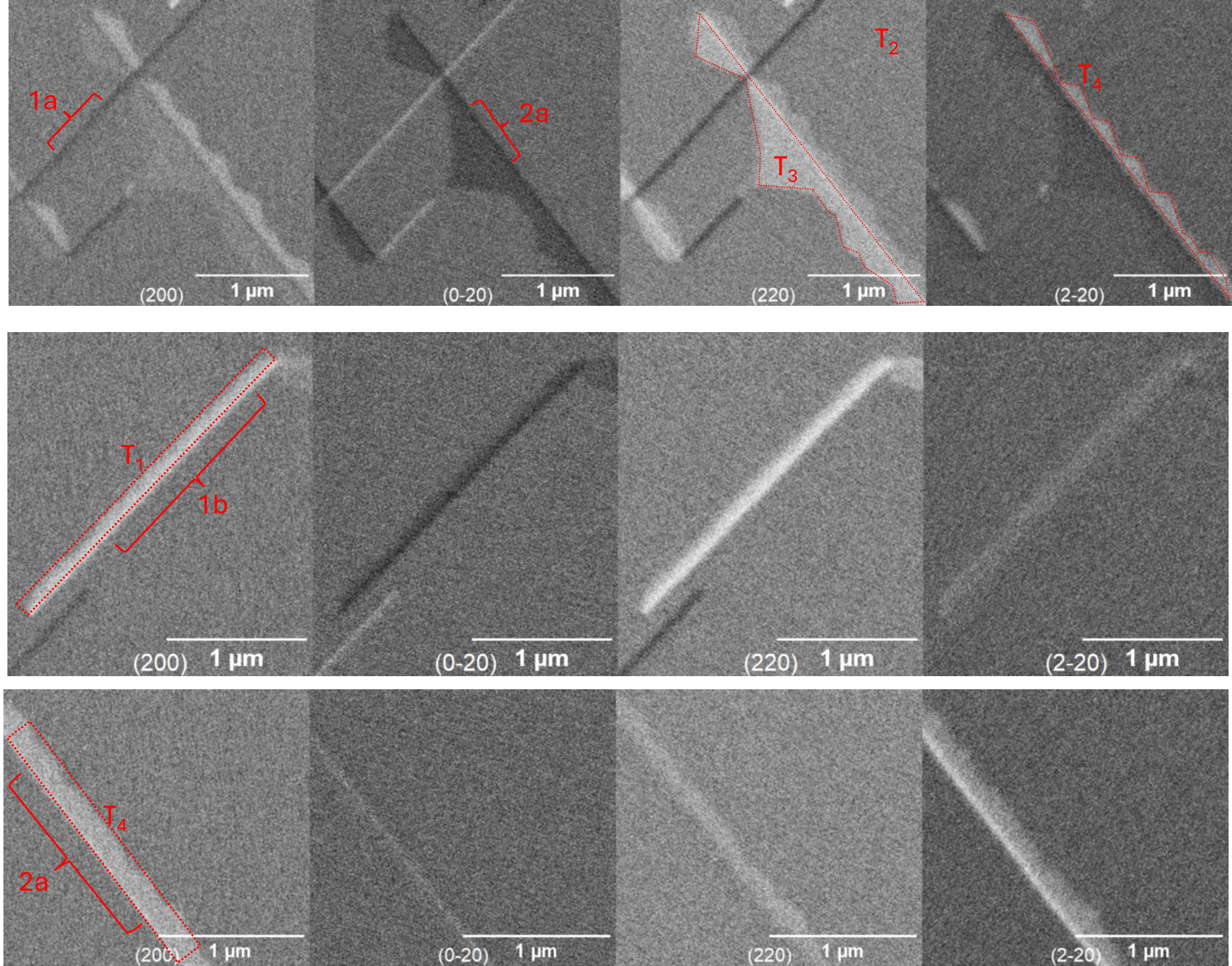}
\caption{Several 60$^{\circ}$ MDs identified by their ECCI sequence. See Table \ref{tab:s4} for details. The observed ECCI contrast sequence allows identification of the T-domains. The location of the flanking domain and its type match the magnetoelastic model for MD stress-field templated T-domains in NiO.}
\label{fig:s8}
\end{figure}

For this 28 nm sample, the SEM-ECCI data provide independent cross-checks of the parent domain assignment and unique child domain identification allowing mapping of the location of all 4 distinct T-domains. Finally, we note that MD 2a or 2a' also show a thinner secondary domain ($d_2$) on the opposite flank from the primary domain $d_3$ identified as $T_3$. Our secondary prediction (case B) is $T_4$, matching the observations. Thus, we find a self-consistent unique assignment of the T-domains forming at 60$^{\circ}$ MDs based on the $E_{int}$ magneto-elastic energy. A few zoomed examples of MDs with T-domain flanks in the 28 nm sample, are shown in Fig. \ref{fig:s8}.

One expects the model only to work in cases where the MDs are long and well-isolated, which is certainly not the case in the 46nm sample where cross-slip is prevalent. However, the same method, as described above, is used to identify $T_p=T_1$, $d_4=T_3$, and $d_5=T_4$ for the 46 nm sample. It is a bit surprising to find that this simple model still accurately predicts the T-domain formation in this highly cross-slipped sample, which emphasizes the importance of the strong local core stress-field of the MDs to this process.

It is tempting to speculate on the nucleation and growth dynamics of the T-domains and antiferromagnmetic domains in general given the results here. Clearly the stress fields from local crystallographic defects, in particular dislocations, serve as highly favorable nucleation sites for T-domains in NiO. Such defects are ubiquitous in crystals, not just epitaxial layers. Given the large magneto-elastic energies involved near the core of a dislocation, there should also be substanatial changes to the local Neel temperature, such that in the vicinity of an edge dislocation, $T_N$ is larger than the surrounding undislocated matrix. Thus, in all real crystals, one expects an initial dislocation to nucleate a T-domain above the global Neel temperature, which then acts as a pre-existing antiferromagnetic domain that can expand into the surrounding crystal. In our films, we imagine the parent T-domain forms in this fashion above the bulk Neel temperature. Due to subtle differences in the stress field due to the MD network, one particular MD nucleates the parent domain first, while the child domains form at a slightly lower temperature.

\section{Dislocation Core Chemistry vs. Long-Range Elastic Strain}

Here, we evaluate whether localized dislocation core defects or long-range elastic stress fields govern the $T$-domain morphology and thermal memory in epitaxial $\text{NiO/MgO}(001)$. Several physical constraints indicate that long-range elasticity is the primary driving mechanism:

(i) The spatial extent of the features and symmetry breaking are indicative of long range stress fields. Chemical, electronic, or uncompensated spin perturbations at the dislocation core  are strictly localized to a radius of $r < 2\text{ nm}$. In contrast, the $T$-domain slabs observed via ECCI extend laterally across $100\text{--}400\text{ nm}$. Localized phenomena, such as core ferromagnetism \cite{sugiyama_ferromagnetic_2013} or charge accumulation, lack the spatial reach to generate the long-range anisotropic fields required to break the fourfold degeneracy of the $T_1\text{--}T_4$ variants across these large volumes.
    
(ii) The elastic shear stress field ($\sigma_{ij}^{\text{MD}}$) of an interfacial misfit dislocation decays slowly as $1/r$, maintaining substantial stress levels ($\sim 10\text{--}100\text{ MPa}$) even at lateral distances of $50\text{--}100\text{ nm}$ from the core. Through Eshelby magnetoelastic coupling, this shear field yields deep interaction energy minima of roughly $-120\text{ meV/nm}^3$ near the core. 
    
(iii) The magnetoelastic driving force exceeds both the intrinsic magnetocrystalline anisotropy and volumetric domain wall penalties ($\sim 0.02\text{--}0.06\text{ meV/nm}^3$) by orders of magnitude. This massive energetic advantage confirms that long-range magnetoelastic shear strain, rather than localized core magnetic moments, fundamentally pins and stabilizes the observed $T$-domain variants.

\bibliographystyle{apsrev4-2}
\bibliography{aapmsamp}

\end{document}